\documentclass[prb,aps,twocolumn,showpacs]{revtex4-1}

\usepackage{times}
\usepackage{graphicx}
\usepackage{float}
\usepackage{latexsym,amsmath,amssymb,bm,euscript}
\usepackage{color}
\usepackage{subfigure}
\usepackage{epstopdf}
\usepackage{type1cm}
\usepackage{ulem}
\fontsize{12pt}{18pt}

\usepackage{appendix}

\begin{document}
\title{Fractional Quantum Hall Bilayers at Half-Filling: Tunneling-driven Non-Abelian Phase}
\author{W. Zhu$^1$, Zhao Liu$^2$, F. D. M. Haldane$^3$ and D. N. Sheng$^1$}
\affiliation{$^1$Department of Physics and Astronomy, California State University, Northridge, California 91330, USA}
\affiliation{$^2$Dahlem Center for Complex Quantum Systems and Institut f\"ur Theoretische Physik, Freie Universit\"at Berlin, Arnimallee 14, 14195 Berlin, Germany}
\affiliation{$^3$Department of Physics, Princeton University, Princeton, NJ 08544, USA}

\begin{abstract}
Multicomponent quantum Hall systems with internal degrees of freedom
provide a fertile ground for the emergence of exotic quantum liquids.
Here we investigate the possibility of non-Abelian topological order in the
half-filled fractional quantum Hall (FQH) bilayer system driven by the tunneling effect between two layers.
By means of the state-of-the-art density-matrix renormalization group, we unveil ``finger print'' evidence
of the non-Abelian Moore-Read Pfaffian state emerging in the intermediate-tunneling regime,
including the ground-state degeneracy on the torus geometry and the
topological entanglement spectroscopy (entanglement spectrum and topological entanglement entropy)
on the spherical geometry, respectively.
Remarkably, the phase transition from the previously identified Abelian $(331)$ Halperin state
to the non-Abelian Moore-Read Pfaffian state is determined to be continuous,
which is signaled by the continuous evolution of the universal part of the entanglement spectrum,
and discontinuities in the excitation gap and the derivative of the ground-state energy.
Our results not only provide a ``proof-of-principle'' demonstration of
realizing a non-Abelian state through coupling different degrees of freedom,
but also open up a possibility in FQH bilayer systems for detecting different chiral $p-$wave pairing states.
\end{abstract}

\date{\today}


\maketitle

\tableofcontents

\section{Introduction}
\begin{figure}[t]
 \begin{minipage}{0.98\linewidth}
 \centering
 \includegraphics[width=0.9\linewidth]{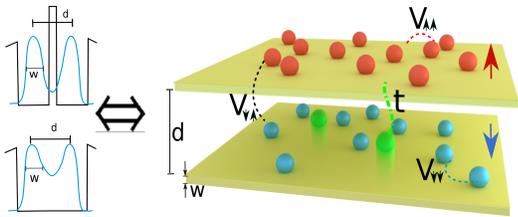}
 \end{minipage}
 \caption{A schematic diagram for the FQH bilayer system. Both double QW and single wide QW systems can be mapped to
 a bilayer system, where electrons interact with each other through intralayer interaction $V_{\sigma \sigma}$ and
 interlayer interaction $V_{\sigma\overline{\sigma}}$, and the electron tunneling $t_\perp$ is tunable between two separated layers.
 } \label{cartoon}
\end{figure}

When two-dimensional electron systems subject to a strong magnetic field, electron-electron interactions
can drive transitions into a series of remarkable quantum states of matter, dubbed as fractional quantum Hall (FQH) effect \cite{Tsui1982,Laughlin1983}.
The FQH effect is an example of the topological state of matter \cite{Wen1990},
providing a spectacular platform for anyonic statistics in two-dimension:
the emergent excitations carry fractionalized quantum numbers and obey Abelian \cite{Laughlin1983} or non-Abelian quantum statistics \cite{Moore,Greiter1991,Read1999a}.
Among them, the non-Abelian FQH effect is expected to form the substrate for topological quantum computation \cite{Nayak2008}, thus is of great importance.
Albeit vigorous research efforts \cite{Willett1987,Pan1999,Radu2008,Dolev2012,Willett2013,Baer2014},
to date convincing experimental evidence of non-Abelian FQH states are still rare, with $\nu=5/2$ and $12/5$
as two prominent examples realized in single-component FQH systems. 
Compared to single-component systems, multicomponent FQH systems with extra degrees of freedom offer
additional tunable parameters and allow the observation of richer quantum phase diagrams \cite{halperin1983,Haldane1988,Wen2000,Read2000,Wen2011}.
The internal degrees of freedom
correspond to realistic experimental circumstances, for example,
layers, subbands or spins in GaAs quantum wells (QWs) \cite{Suen1992,Eisenstein1992,Suen1994,Shabani2013,Jim2014},
spins or valleys in graphene or AlAs,
which lead to effective multilayers separated by layer distance $d$ with electrons' tunneling $t_{\perp}$ between layers (Fig.~\ref{cartoon}).
Two most notable examples of the multicomponent FQH effects are the observation of quantized Hall plateaus
at total filling factors $\nu_T=1/2$ and $\nu_T=1$ in double QW and wide QW systems.
The $\nu_T=1$ state \cite{Jim2014} is believed to favor a symmetry-breaking state with spontaneous interlayer phase coherence,
which induces a remarkable exciton condensation.
The $\nu_T=1/2$ state \cite{Suen1992,Eisenstein1992,Suen1994,Shabani2013} has turned out to be more interesting and controversial,
as it can be an Abelian Halperin FQH state, but also be a possible platform for realizing non-Abelian anyonic statistics, which has been pursued persistently in the past.\cite{Greiter1992,Ho1995,Nayak1996,Fradkin1999,Read2000,Wen2000,Wen2011,Cappelli2001,Naud2000,Seidel2008,Regnault2008}

In this paper, we focus on the two-component FQH system at total filling $\nu_T=1/2$.
As illustrated in Fig.~\ref{cartoon}, we study
a realistic Hamiltonian containing essential information relevant to experiments:
\begin{widetext}
\begin{eqnarray}
H=\sum_{i<j}^{N_e} [V_{\uparrow \uparrow}(|\mathbf{r}_{i\uparrow}-\mathbf{r}_{j\uparrow}|) +
V_{\downarrow\downarrow}(|\mathbf{r}_{i\downarrow}-\mathbf{r}_{j\downarrow}|) ]
+ \sum_{i,j}^{N_e} V_{\uparrow\downarrow}(|\mathbf{r}_{i\uparrow}-\mathbf{r}_{j\downarrow}|) + H_t \;,
\label{eq:ham1}
\end{eqnarray}
\end{widetext}
where we label two layers by index $\sigma\in \{\uparrow,\downarrow\}$ and
the position of the $i$-th electron in layer $\sigma$ by $\mathbf{r}_{i\sigma}$.
$V_{\uparrow\uparrow}(r)=V_{\downarrow\downarrow}(r)$ is the Coulomb potential
in a single layer with finite width $w$, and
$V_{\uparrow\downarrow}(r)$ is the interlayer
Coulomb interaction incorporating finite interlayer separation $d$.
$H_t$ describes the electron tunneling between two layers and the tunneling strength is $t_{\perp}$ (Details see Appendix \ref{app:method}).
We use the magnetic length $\ell$ and the Coulomb energy $e^2/\ell$ as the units of length and energy respectively throughout this work, where $-e$ is the electron charge.

In limits of the spatial separation $d\rightarrow 0$ and $d\rightarrow \infty$, the bilayer ground states at $\nu_T=1/2$ are compressible
because either the single-layer limit ($d\rightarrow 0$) or the two-isolated-layer limit ($d\rightarrow\infty$)
can be well understood by the ``composite Fermi liquid'' (CFL) theory\cite{Halperin1994}.
In the intermediate regime $d\sim 2-6$, the incompressibility of the system is observed
in various experiments \cite{Suen1992,Eisenstein1992,Suen1994,Shabani2013}.
However, its precise origin is still a long-standing subject.
Numerical simulations \cite{Chakraborty1987,Yoshioka1989,SHe1993,Peterson2010a}
have confirmed that the Abelian $(331)$ Halperin state dominates at vanishing interlayer tunneling ($t_{\perp}=0$) \cite{halperin1983}.
Remarkably, through uncovering the underlying pairing nature of the $(331)$ Halperin state \cite{Greiter1992,Ho1995},
it has been suggested that the tunneling effect may drive the system into a non-Abelian phase \cite{Nayak1996,Fradkin1999,Read2000,Wen2000,Wen2011,Cappelli2001,Naud2000,Seidel2008,Regnault2008},
which  motivates intensive efforts \cite{Nomura2004,Papic2009,Papic2010,Peterson2010a,Park2015} to establish  its existence.
However, previous numerical studies, primarily utilizing exact diagonalization on small system sizes,
are still too limited to reach a consensus. For instance,
the only evidence of a non-Abelian phase was obtained by simply comparing the Coulomb ground state with the
trial wavefunction \cite{Papic2009}.
On the contrary, subsequent studies even suggest that
the CFL \cite{Papic2010} and the $(331)$ Halperin state \cite{Peterson2010a} may still dominate at finite tunneling.
Taken as a whole, to date, the possibility of realizing a non-Abelian state through coupling different degrees of freedom or
tuning experimental relevant interactions remains elusive for the $\nu_T=1/2$ bilayer system,
which urgently calls for revisiting this problem using state-of-the-art techniques \cite{White,Shibata,Feiguin2008,JizeZhao2011,Zaletel2015}.

In this article, we uncover the nature of quantum states and determine the phase diagram
for the FQH bilayer system at $\nu_T=1/2$ by means of
large-scale density-matrix renormalization group (DMRG)\cite{White,Shibata,Feiguin2008,JizeZhao2011,Zaletel2015}
and exact diagonalization (ED) calculations.
The system turns out to host  two different incompressible liquid phases:
one is the Abelian $(331)$ Halperin state, and the other is the non-Abelian Moore-Read (MR) Pfaffian state,
as identified by their different ground-state degeneracies on torus geometry. Remarkably, we demonstrate that they share the same
topological entanglement entropy, but have different characteristic entanglement spectra on the spherical geometry.
Furthermore, we identify a continuous phase transition between these two FQH phases driven by varying the tunneling strength $t_{\perp}$,
reflected by the smooth evolution of the ground-state energy, and discontinuities of the excitation gap and the derivative of the ground-state energy.
Intriguingly, our fingerprint evidence leads to two conclusions related to existing theories and experiments. First,
the MR Pfaffian state can indeed be obtained by coupling different degrees of freedom.
Although such a possibility was predicted about 20 years ago \cite{Greiter1992,Ho1995,Nayak1996,Fradkin1999,Read2000,Wen2000,Wen2011},
convincing and comprehensive evidence directly from a microscopic description was missed until our work.
Second, the previously found FQH $\nu_T=1/2$ plateau in single wide QW experiments \cite{Suen1992,Suen1994,Shabani2013}, where the tunneling strength
is taken to be considerable, \textit{is most likely to} be captured by the MR Pfaffian state and in favor of a nontrivial $p_x+ ip_y$ pairing mechanism.
By reducing the effective tunneling (through tuning electron density),
the system undergoes a transition from  the non-Abelian MR Pfaffian to the weak $p$-wave pairing $(331)$ Halperin state,
while the Hall conductance keeps unchanged.
Thus, the $\nu_T=1/2$ bilayer system provides a promising platform for
realizing  different $p_x+ ip_y$ pairing physics through coupling different degrees of freedom \cite{Teo2014,Vaezi2014,YiZhang2014}
within experimentally attainable parameters.
We believe our work paves the way for future research realizing new classes of non-Abelian states in realistic bilayer systems.
Specific measurements for identifying the bilayer non-Abelian state in experiments are also discussed.

\section{Energy Spectrum}
We first investigate the torus geometry with periodic boundary condition,
where different topological states can be distinguished by their ground-state degeneracies. At
filling factor $\nu_T=1/2$, apart from a two-fold degeneracy coming from the center-of-mass motion,
there can be additional degeneracy occurring due to the multicomponent or the topological nature of the state, which is four-fold
for the $(331)$ Halperin state, and three-fold for the MR Pfaffian state\cite{Greiter1991}.
Here we will inspect the low-energy spectrum as a function of the tunneling strength using DMRG.

\begin{figure}[t]
 \begin{minipage}{0.99\linewidth}
 \centering
 \includegraphics[width=0.9\textwidth]{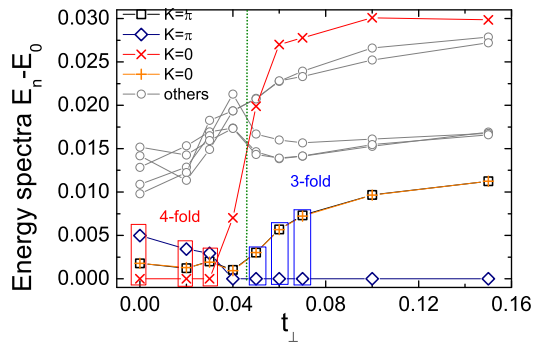}
 \end{minipage}
 \caption{(a) Energy spectra as a function of the tunneling strength $t_{\perp}$, obtained on a square torus with $N_e=12$ electrons by DMRG.
 Different momentum sectors are labeled by different symbols. We highlight the ground-state degeneracy by boxes. Here we only show two lowest energy levels in $K=0$ (red cross, orange star) and $K=\pi$ (black square, navy diamond) sectors, and one lowest energy level in other momentum sectors. Due to the $C_4$ symmetry on the square torus, one ground state in the $K=0$ sector (orange star) has the nearly the same energy with one in the $K=\pi$ sector (black square). The dashed line marks the level crossing between the ground state in the $K=0$ sector (red cross) and high excited states, indicating a quantum phase transition. All calculations are performed at layer distance $d=3.0$ and layer width $w=1.5$.
 } \label{torus}
\end{figure}

In Fig.~\ref{torus}(a), we show the results
for $N_e=12$ electrons obtained by DMRG at layer width $w=1.5$ and layer distance $d=3.0$, where the degeneracy due to the center-of-mass motion has been excluded. When the tunneling is weak ($t_\perp<0.04$),
we identify the multiplet of four ground states in the  spectrum as a signal of the $(331)$ Halperin state.
With the increasing of the tunneling strength, the four-fold ground-state degeneracy is gradually destroyed.
One state with momentum $K=0$ (marked as red cross) is being gapped out
for sufficiently large $t_{\perp}$, leaving other three states in the ground-state manifold.
Importantly, we find a region ($t_{\perp}>0.04$) where the correct three-fold MR Pfaffian degeneracy is visible,
despite a finite energy splitting among the three ground states.
Here we would like to point out, working on the larger system sizes is the key to
reach this exciting result. In the system size $N_e<12$, one energy state from
$K\neq 0, \pi$ comes down and eventually forms a gapless branch in the low energy spectrum (Appendix Sec. \ref{app:ED}),
which prevents previous work \cite{Papic2010} from reaching a positive conclusion of the three-fold MR Pfaffian degeneracy.

\section{Entanglement Spectroscopy}
To uncover the topological nature of different quantum phases,
we move to the spherical geometry, and perform the entanglement-based diagnosis.
This geometry is commonly used for accessing larger systems as the unique ground state (selected by the finite-size shift, see Appendix \ref{app:method}) on the sphere facilitates the computation task. We analyze the topological entanglement entropy (TEE) \cite{Kitaev2006,Levin2006}
and the entanglement spectrum (ES)\cite{Haldane2008} in different tunneling regimes with different ground-state degeneracies on the torus, and demonstrate that they accurately match the predictions
for the $(331)$ Halperin state and the MR Pfaffian state in the weak- and intermediate-tunneling regime, respectively.
Importantly, all characterizations of phases are robust and stable for various system sizes [from $N_e=14$ to $24$
(see Appendix Sec. \ref{app:ES})].

\subsection{Topological Entanglement Entropy}
The entanglement entropy of a bipartite quantum state $|\Psi\rangle_{AB}$ is defined as $S_A=-{\rm Tr}\rho_A\ln\rho_A$, where
$\rho_A={\rm Tr}_B(|\Psi\rangle \langle\Psi|)$ is the reduced density matrix of the subsystem $A$.
For a gapped topological order in two-dimension, the area law $S_A=\alpha |\partial A|-\gamma$ holds, where $|\partial A|$ is the boundary length of the subsystem $A$,
and the TEE $\gamma$ is related to the total quantum dimension
$\mathcal{D}$ by $\gamma=\ln \mathcal{D}$ \cite{Kitaev2006,Levin2006}.
Since $\mathcal{D}$ contains the information about quasiparticles, the TEE can determine whether a topological phase belongs to
the universality class of a given topological field theory.

We make two identical single cuts, each applied to one sphere in our bilayer system, to divide all Landau level orbitals into two parts. The subsystem $A$ contains $2l_A$ Landau level orbitals in total ($l_A$ consecutive orbitals in each northern hemisphere). For partitions with different $l_A$, since the boundary length of the cut is proportional to $\sqrt{l_A}$, we expect the area law $S_A(l_A)=\alpha\sqrt{l_A}-\gamma$.
Figs.~\ref{ES}(a) and \ref{ES}(d) show the numerically calculated orbital-cut entanglement entropy $S_A(l_A)$
as a function of $\sqrt{l_A}$ for tunneling strength $t_{\perp}=0.03$ and $0.10$ at layer width $w=1.5$ and layer distance $d=3.0$.
First of all, the approximately linear part of $S(l_A)$ shows
a negative intercept in the limit of $l_A \rightarrow 0$, indicating a nonzero TEE.
Through the finite-size scaling (red line) based on the raw data of $N_e=22$, we
extract the TEE as $\gamma\approx1.119\pm 0.143$ and $\gamma\approx1.031\pm0.074$
for $t_{\perp}=0.03$ and $t_{\perp}=0.10$, respectively.
Interestingly, the $(331)$ Halperin state and the MR Pfaffian state share the same theoretical value of TEE -- they have the same
total quantum dimension $\mathcal{D}=\sqrt{8}$, despite hosting different types of quasiparticles
[the $(331)$ Halperin state hosts $8$ different Abelian quasiparticles, while the MR Pfaffian has $4$ Abelian and $2$ non-Abelian quasiparticles].
Indeed, both of our extracted results are very close  to each other,
in agreement with the expectation $\gamma=\ln \sqrt{8}\approx1.037$ (blue dashed line).
Although the definite Abelian or non-Abelian nature cannot be determined by TEE,
the observation of a nonzero TEE signals the topologically non-trivial state in the finite-tunneling regime.

\subsection{Orbital Entanglement Spectrum}
The orbital ES, defined as the spectrum of $-\ln\rho_A$, encodes the information of edge excitations \cite{Haldane2008,Wen1995} and has been widely
used to identify the emergent FQH phase in a microscopic Hamiltonian \cite{Haldane2008,WZhu2015,Mong2015}.
For various single-layer FQH states, including the
Laughlin, Moore-Read \cite{Haldane2008} and Read-Rezayi states \cite{WZhu2015,Mong2015},
the ES has a universal low-energy structure mimicing the pertinent edge excitation spectrum, 
which is separated from the high-energy non-universal part by a finite ES gap.
In our bilayer $\nu_T=1/2$ system, different candidates host distinct edge excitations, so
we anticipate to distinguish them by the orbital-cut ES.
Very recently, the orbital ES diagnosis was also extended to bilayer $1/3+1/3$ systems \cite{Zhao2015,Scott2015},
albeit there is no signal of non-Abelian states in such systems with pure Coulomb interaction (without artificially tuning
pseudopotential parameters).

Edge excitations of a specific FQH state are characterized by the degeneracy pattern of the spectrum when plotted versus appropriate quantum numbers, for example, the angular momentum $L_z$ on the sphere. The edge of the $(331)$ Halperin state can be described by two chiral boson fields (Appendix \ref{app:theory} ),
thus the corresponding edge excitation spectrum exhibits degeneracy in angular momentum sectors $\Delta L_z=0,1,2,3,\cdots$ as
(Appendix \ref{app:counting})
\begin{eqnarray*}
   &&\textrm{even}: 1,2,7,14,\cdots, \\
   &&\textrm{odd}: 2,4,10,20,\cdots,
\end{eqnarray*}
where two sequences are distinguished by the even (odd) number of electrons, and $\Delta L_z=L_z-L_{z,\min}$ with $L_{z,\min}$ the angular momentum where no edge excitations occur.
The edge excitations of the MR Pfaffian state, composed of a Majorana fermion mode
and a charged boson mode (Appendix \ref{app:theory}), should follow the degeneracy pattern
(Appendix \ref{app:counting})
\begin{eqnarray*}
   &&\textrm{even}: 1,1,3,5,\cdots, \\
   &&\textrm{odd}: 1,2,4,7,\cdots.
\end{eqnarray*}
In contrast, the CFL state does not develop a gapless ``edge''  spectrum  separated from other spectrum by a gap  due to its compressible nature.

\begin{figure}[!htb]
 \begin{minipage}{0.99\linewidth}
 \centering
 \includegraphics[width=0.30\linewidth]{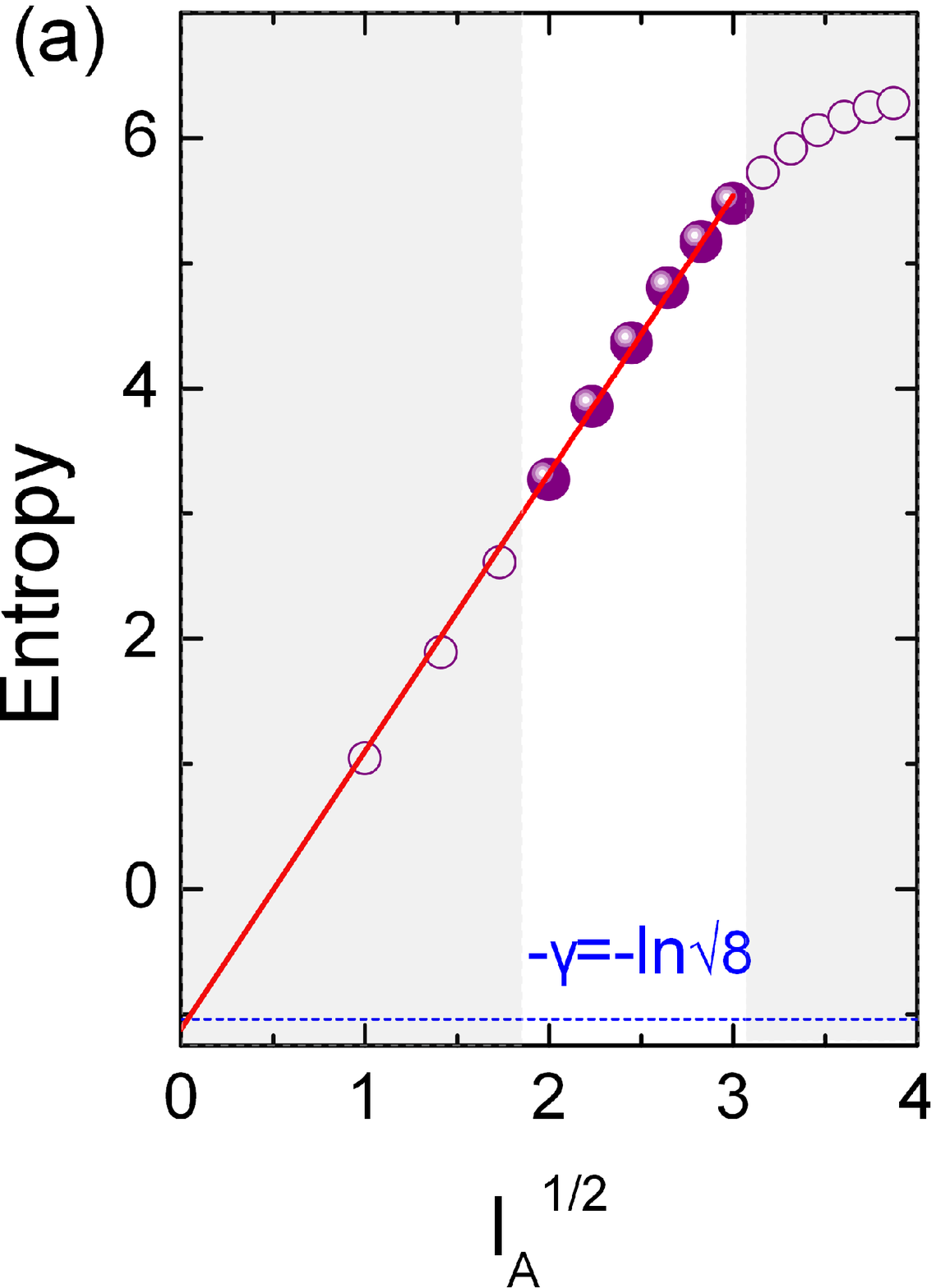}
 \includegraphics[width=0.62\linewidth]{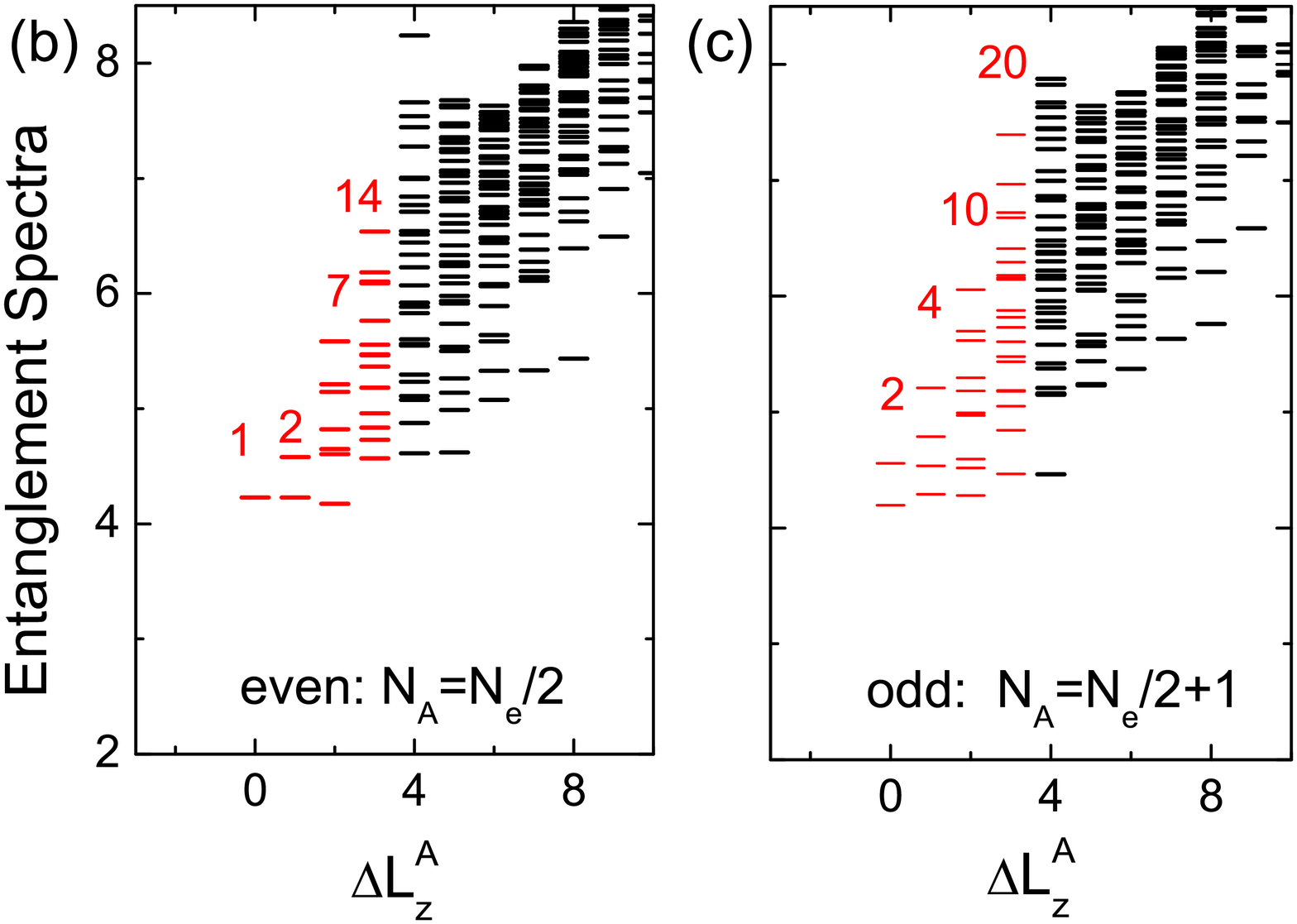}
 \includegraphics[width=0.30\linewidth]{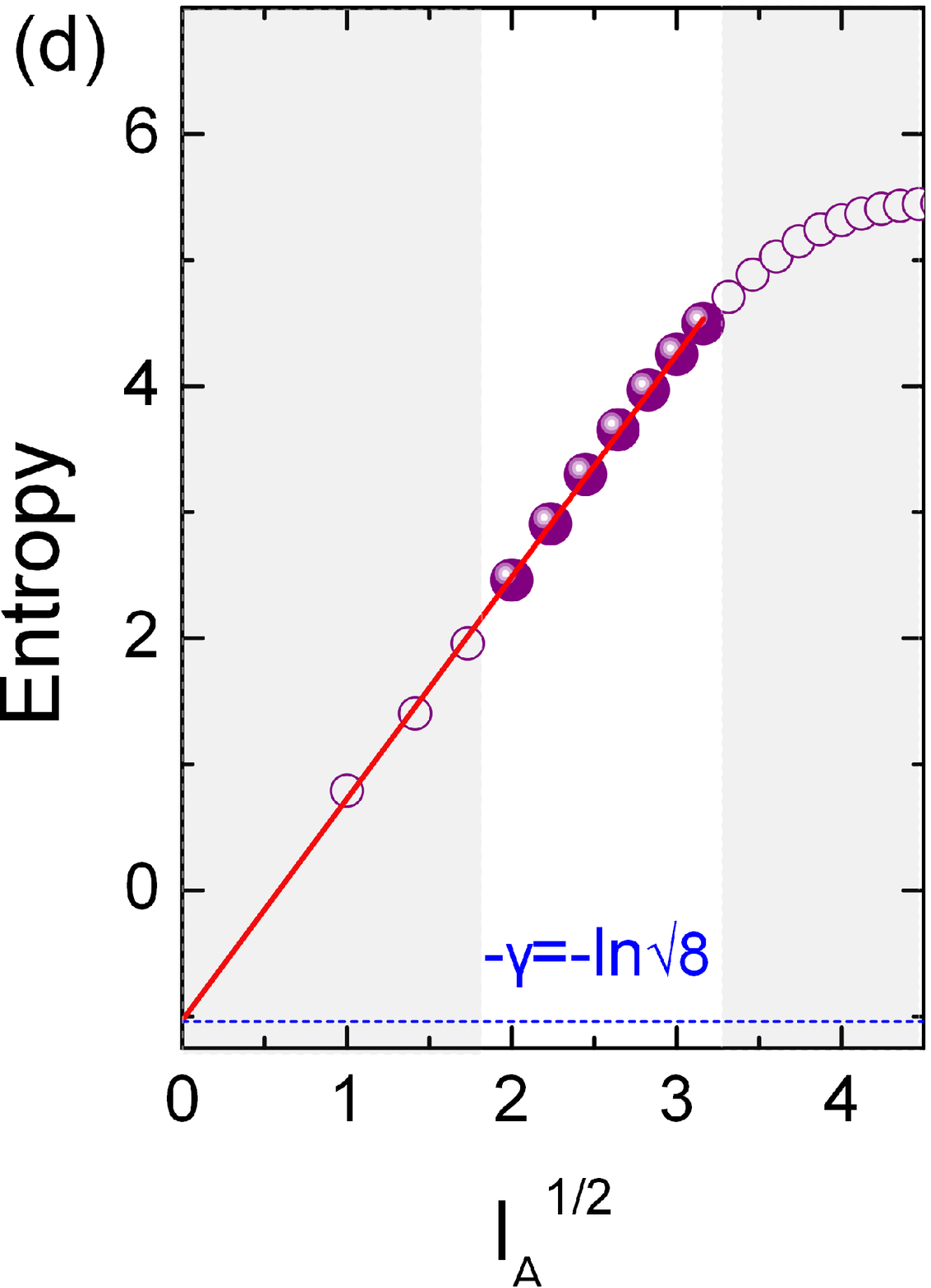}
 \includegraphics[width=0.62\linewidth]{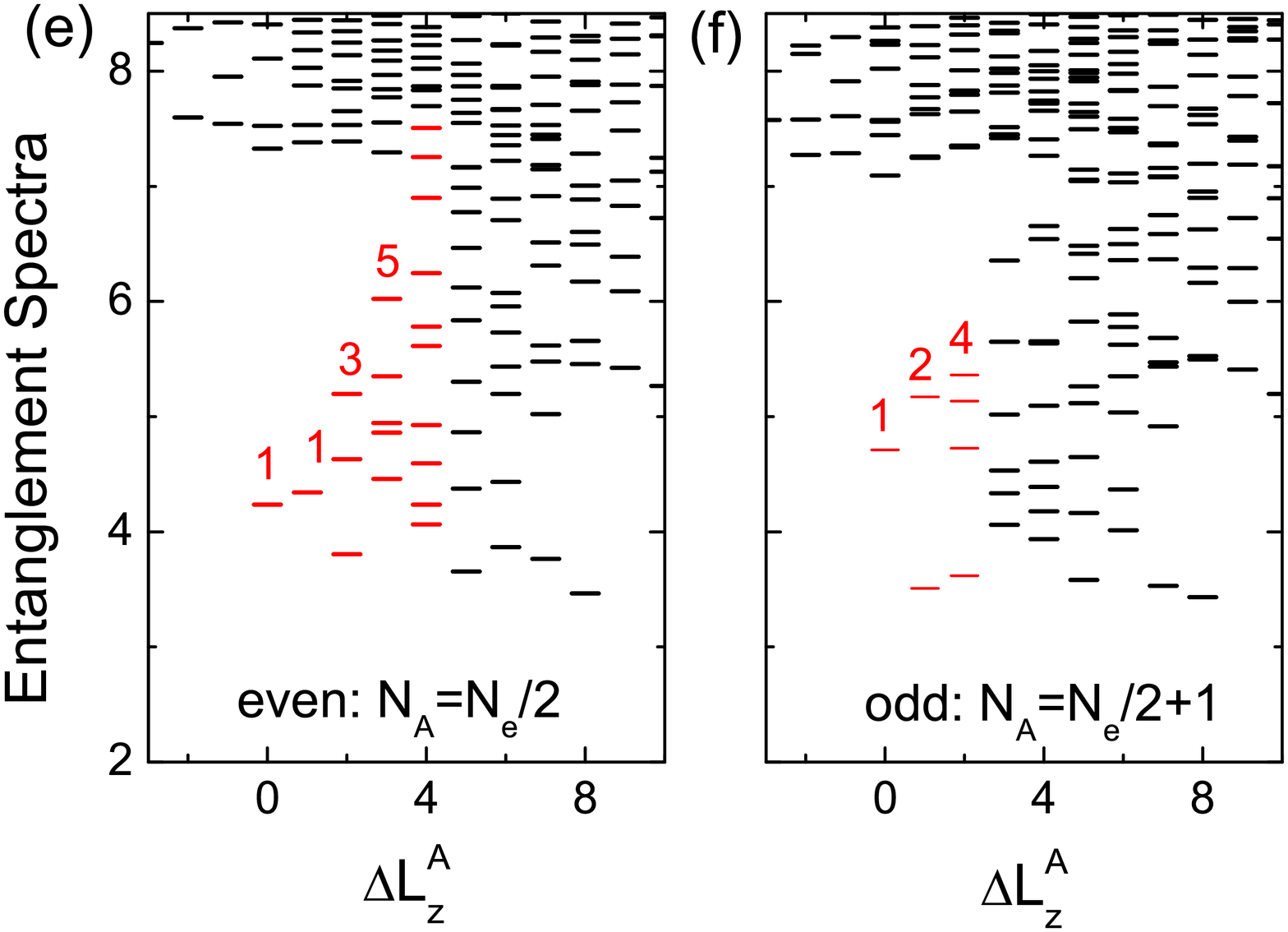}
 \end{minipage}
 \caption{(a,d) The entanglement entropy $S(l_A)$ for (a) the $(331)$ Halperin state and (d) the MR Pfaffian state
 as a function of $\sqrt{l_A}$.
The open circles were discarded in the extrapolation because they either represent
very small subsystems violating the area law or suffer from the finite-size saturation effect (shaded by grey).
The linear extrapolated $\gamma$ in both cases are in agreement with the predicted value $\ln \sqrt{8}$ (blue dashed line).
 (b,c,e,f) The low-lying orbital ES of (b,c) the $(331)$ Halperin state and (e,f) the MR Pfaffian state,
 with even or odd electrons in the half-cut subsystem.
 The countings matching the degeneracy patterns given in the text are labeled by red.
 All calculations are performed at system size $N_e=22$ with layer width $w=1.5$, layer distance $d=3.0$,
 and tunneling strengthes (a-c) $t_{\perp}=0.03$ for  the $(331)$ state and (d-f) $t_\perp=0.10$ for  the MR Pfaffian state.} \label{ES}
\end{figure}

In Figs.~\ref{ES}(b,c) and \ref{ES}(e,f), we show the DMRG obtained ES for $t_{\perp}=0.03$ and $0.10$ at layer width $w=1.5$ and layer distance $d=3.0$.
At weak tunneling $t_\perp=0.03$, we find that the low-lying ES levels exactly match the degeneracy patterns of the $(331)$ edge spectrum in
the first four $\Delta L_z^A$ sectors, i.e., $1,2,7,14$ for even $N_A$ and $2,4,10,20$ for odd $N_A$, where $N_A$ and $\Delta L_z^A$ are the number of electrons and the angular momentum in the subsystem $A$, respectively.
Those low-lying levels are separated from higher ones by a large ``entanglement gap''.
At stronger tunneling $t_\perp=0.1$, the low-energy ES
clearly displays the degeneracy patterns of the MR Pfaffian edge spectrum, i.e., $1,1,3,5$ for even $N_A$ and $1,2,4$ for odd $N_A$.
Different low-lying ES structures provide compelling evidence that the ground state undergoes a transition from the $(331)$ Halperin phase
to the MR Pfaffian phase by tuning $t_\perp$.
As shown in Fig. \ref{transition} (a), with the increase of tunneling $t_{\perp}$, some ES levels belong to the $(331)$ Halperin state can be continuously gapped out.
After a new entanglement gap $\Delta_1$ is well-developed ($t_\perp>0.05$), the desired ES structure for MR Pfaffian state appears,
perfectly matching the prediction that one branch of Majarona fermion mode can be continuously gapped out by the tunneling effect (Appendix Sec. \ref{app:theory}).

\section{Quantum Phase Transition and Phase Diagram}
\begin{figure}[t]
 \begin{minipage}{0.98\linewidth}
 \centering
 \includegraphics[width=0.99\textwidth]{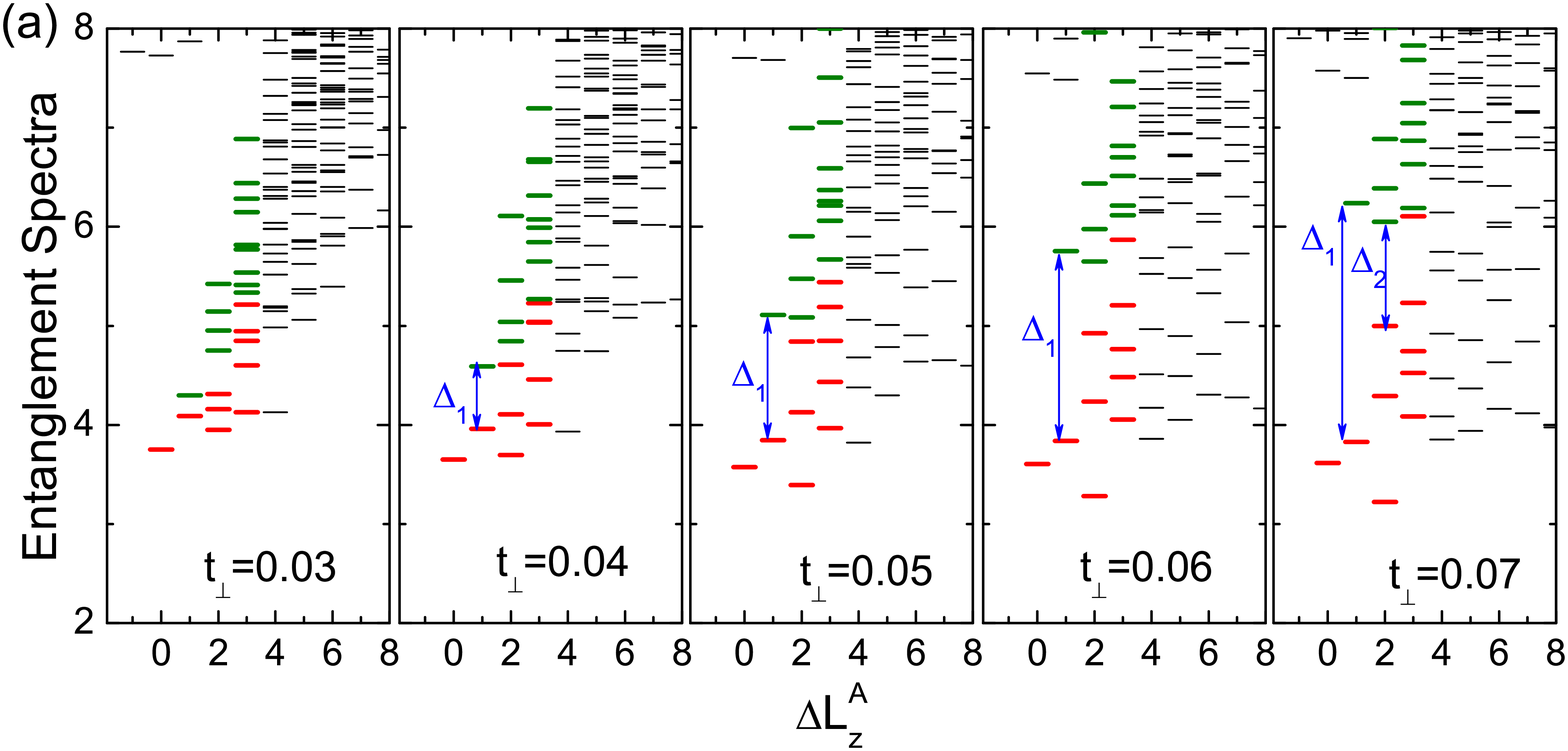}
 \includegraphics[width=0.49\textwidth]{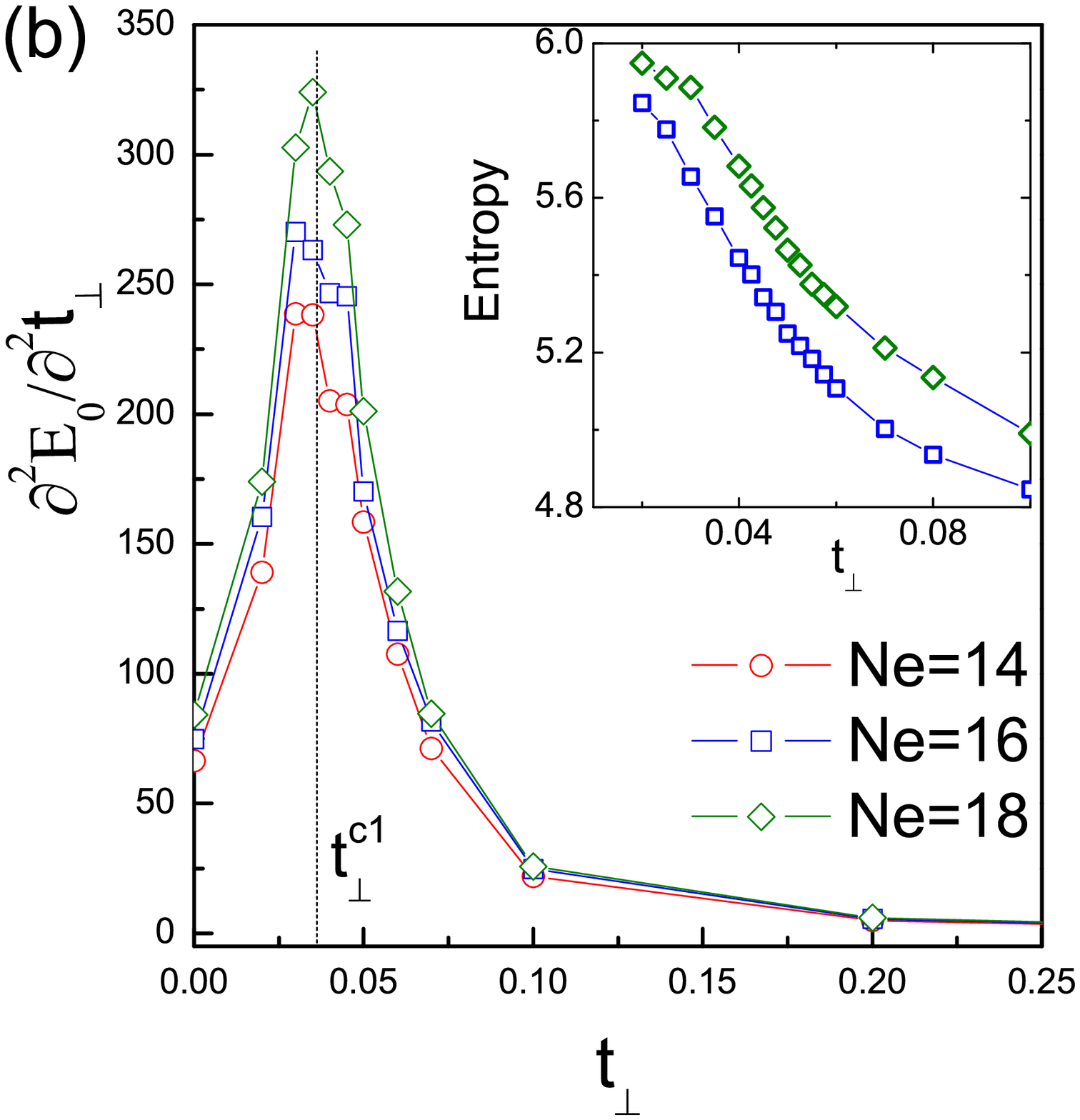}
 \includegraphics[width=0.49\textwidth]{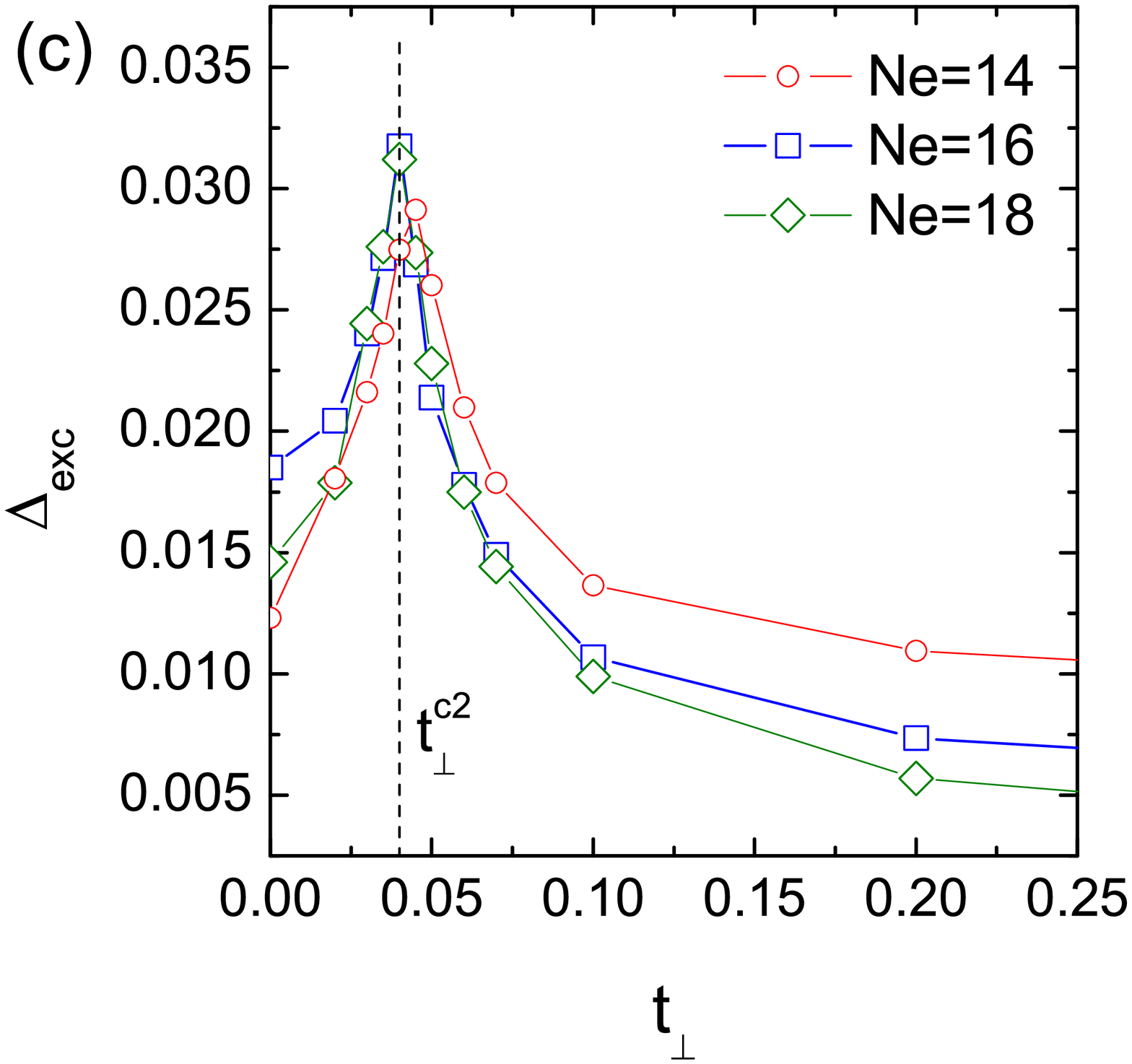}
 \end{minipage}
 \caption{The continuous phase transition from the $(331)$ Halperin state to the MR Pfaffian state as a function of $t_\perp$ on the sphere.
 (a) Evolution of ES versus $t_\perp$. The expected levels for the MR Pfaffian state are labeled by red,
 and redundant levels originally from the $(331)$ Halperin state are labeled by green. $\Delta_{\Delta L_z^A}$ measures the entanglement gap
 of the MR Pfaffian state in the $\Delta L_z^A$ sector. We consider the half-cut subsystem with even number of electrons for $N_e=18$.
 (b) Partial derivative $\partial^2 E_0/\partial^2 t_{\perp}$ as a function of $t_{\perp}$ for different system sizes
 $N_e=14$ (red), $16$ (blue) and $18$ (green).
 Inset: Evolution of the ground-state entropy with $t_{\perp}$.
 (c) The excitation gap $\Delta_{\rm{exc}}$ as a function of $t_{\perp}$ for various system sizes.
All calculations are performed at layer width $w=1.5$ and layer distance $d=3.0$.}
 \label{transition}
\end{figure}

To uncover the nature of the quantum phase transition driven by $t_{\perp}$, we study the evolutions
of the ground state and the lowest excited state on the spherical geometry, which have different total angular momenta $L_z$.
We choose fixed layer width $w=1.5$ and layer distance $d=3.0$ in these calculations.
In Fig.~\ref{transition}(b), we first investigate how the ground-state energy $E_0$ varies with $t_{\perp}$.
Although $E_0$ smoothly changes with $t_{\perp}$,  we find
a discontinuity in $\partial^2 E_0/\partial^2 t_{\perp}$ around $t^{c1}_{\perp}\approx 0.037$.
The singularity becomes sharper by increasing the system size, indicating a second-order phase transition in the thermodynamic limit.
In Fig.~\ref{transition}(c), we show the excitation gap as a function of $t_{\perp}$,
defined as the energy difference between the first excited
state and the ground state [$\Delta_{\rm{exc}} = E_1(L_z\neq 0)-E_0(L_z=0)$].
$\Delta_{\rm{exc}}$ remains finite for all $t_{\perp}$,
consistent with the incompressible nature of the ground state.
Interestingly, the excitation gap develops a peak around $t^{c2}_{\perp}\approx 0.04$.
This upward cusp is related to
a level-crossing between the lowest excited state and higher energy levels \cite{Nomura2004,Peterson2010a}.
These observations lead to two remarks here.
First, our calculations support that the transition detected by the ground state and the lowest excited state occurs almost simultaneously
($t^{c1}_{\perp}\approx t^{c2}_{\perp}$).
Second, our results indicate that the ground state evolves continuously from the $(331)$ Halperin phase to the MR Pfaffian phase,
while the excited state with quasihole or quasiparticle excitations changes discontinuously near the phase boundary.

Compared with the spherical geometry with zero genus where we can only reach one topological sector related to the ``highest density profile''
(see Appendix Secs. \ref{app:counting} and \ref{app:theory2}), the torus geometry with access to all topological sectors
provides a full picture of the gap closing and the continuous phase transition.
As shown in Fig.~\ref{torus}, the energy gap relative to the $(331)$ manifold closes around $t = t^c_\perp$ with one $K=0$ state in the $(331)$ manifold
being continuously gapped out without any level crossing in the low-energy spectrum (also see discussion below).
To sum up, our findings provide evidence of
the continuous transformation between two triplet pairing states, which was predicted 20 years ago\cite{Greiter1992,Ho1995,Wen2000,Read2000}.

Intriguingly, the continuous phase transition between the $(331)$ Halperin state and the MR Pfaffian state
can be understood\cite{Wen2000} from several perspectives. 
In Appendix Sec.~\ref{app:theory}, in addition to the wave-function equivalence,
we propose two independent perspectives to understand the transition in the bulk and on the edge, respectively.
First, by the perturbation theory, we
construct a low-energy effective model, which clearly shows that, at least in the thin-torus limit\cite{Seidel2008,Vaezi2014}, the system can indeed undergo a continuous phase transition (with the same critical behavior as the transverse field Ising model \cite{Sachdev2000}) when the tunneling $t_{\perp}$ increases, and one state in the ground-state manifold is gapped out, thus changing the ground-state degeneracy from the $(331)$ type to the MR Pfaffian type. We believe this conclusion is still true when the system adiabatically deforms from the thin-torus limit to the square torus.
Second, starting from the edge theory of the $(331)$ Halperin state described by two chiral bosons (with total central
charge $c = 2$) \cite{Wen1990c,Wen1992}, we find that the interlayer tunneling tends to produce a Majorana neutral mode
carrying $c = 1/2$ in addition to the usual $c = 1$ bosonic charge mode \cite{Naud2000,Fradkin_Book}, thus reaching the edge theory of the MR Pfaffian state.
Therefore, from the viewpoints of the wave-function equivalence, the bulk theory in the thin-torus limit,
and the effective edge theory, a continuous phase transition is allowed between these two triplet pairing FQH states \cite{Ho1995,Read2000}.

\begin{figure}[t]
 \begin{minipage}{0.98\linewidth}
 \centering
 \includegraphics[width=0.5\linewidth]{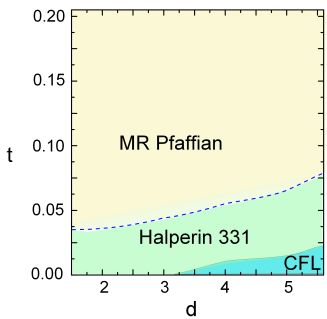}
 \end{minipage}
 \caption{
 The phase diagram of the FQH bilayer at $\nu_T=1/2$ in terms of the layer distance $d$ and tunnelling strength $t_{\perp}$,
 obtained from $N_e=18$ on the sphere with layer width $w=1.5$.  The continuous phase transition from the $(331)$ Halperin state to the MR Pfaffian state is labeled by dashed line, while
the solid line marks the transition between the $(331)$ Halperin state and a possible CFL.
 } \label{phase}
\end{figure}

At last, we present a quantum phase diagram for the FQH bilayer system at $\nu_T=1/2$,
as functions of experimentally relevant parameters $d$ and $t_{\perp}$ in Fig.~\ref{phase}.
Different phases and their phase boundaries are determined by the entanglement spectrum based on the $N_e=18$ data on the sphere.
We find three different phases: the $(331)$ Halperin phase, the MR Pfaffian phase,
and the compressible CFL phase.
When $t_{\perp}$ is small and $d$ is relatively large,
two layers are effectively decoupled with each at $\nu=1/4$ (in the $d\rightarrow\infty$ limit) and the ground state is a well-known CFL.
At small $t_{\perp}$, the ground state
is in the $(331)$ Halperin phase, then a phase transition to the MR Pfaffian state occurs at $t_{\perp}\sim 0.04-0.07$ (depending on the value of $d$).
The intermediate tunneling regime $t_{\perp}\sim 0.05-0.1$ has larger excitation gap as shown in Fig.~\ref{transition}(c), where the MR Pfaffian state is most likely to be observed experimentally (Appendix Sec.~\ref{app:exp}).
Interestingly, the maximal excitation gap in the intermediate tunneling regime [Fig.~\ref{transition}(c)]
qualitatively agrees with the experimental observation \cite{Suen1994}.
This also supports that MR Pfaffian state is more robust in the intermediate-tunneling regime while the $(331)$ Halperin state is stable in the weak-tunneling regime.
In addition, we point out that, even though the MR Pfaffian phase is shown to be remarkably robust in the intermediate-tunneling regime,
we are less certain about the fate of the state in the strong-tunneling limit ($t_\perp\rightarrow \infty$) (see Appendix Sec.~\ref{app:ED2})
due to other competing phases.

\section{Conclusion}
In this work, we use density-matrix renormalization group and exact diagonalization
techniques to study a fractional quantum Hall (FQH) bilayer system at total half-filling.
In the phase diagram in terms of the experimentally accessible parameters (layer separation $d$,
interlayer tunneling $t_{\perp}$, and layer width $w$),
we find two different incompressible phases:
the Abelian $(331)$ Halperin phase in the weak-tunneling regime and the non-Abelian Moore-Read (MR) Pfaffian state for the intermediate tunneling
strength, as identified by the ground-state degeneracy on the torus geometry and the topological entanglement-based diagnosis on the spherical geometry.
The results on different geometries are consistent with each other and give similar phase boundaries.
We also establish that the transition between these two phases is continuous,
which verifies the theoretical conjecture that a continuous phase transition is allowed between two triplet pairing states\cite{Ho1995,Read2000}
with distinct quasi-particle excitations.
Our work clearly demonstrates that for realistic two-component FQH systems, the
non-Abelian MR Pfaffian state is indeed a stronger candidate than the Abelian $(331)$ Halperin state
in the intermediate-tunneling regime.

We believe that our work will motivate experimental activities searching for the non-Abelian phase in bilayer structures at total half-filling.
Some existing theoretical proposals can be used to identify the corresponding edge physics experimentally\cite{Wen1995,Bishara2009,Fiete2008}.
For example, the quasiparticle tunneling conductance acrossing quantum point contacts
allows the extraction of the dimensionless interaction parameter $g$, which reflects
the topological order in the bulk and can be directly compared with the theoretical
expectations of $g=1/4$ for the MR Pfaffian state and $g=3/8$ for the $(331)$ Halperin state \cite{Radu2008,Baer2014}.
Another approach is to probe the edge density fluctuation
when the sample is coupled to a nearby quantum dot \cite{Fiete2008}.
Furthermore, the measurement of drag Hall conductance in double QWs can be performed to identify different phases. The $(331)$ Halperin state has the quantized Hall drag conductance, while the MR Pfaffian state has a strong density fluctuation with non-quantized Hall drag conductance (see Appendix \ref{app:exp}).
On the theoretical sides, our work, with combined DMRG and ED methods, also paves the way for future studies of other
multicomponent systems with the aim to search for more exotic FQH states.

W.~Z. thanks Z.-X. Hu, M. Peterson, C. J. Wang and T. S. Zeng for fruitful discussion,
and L. Wang for preparing Fig.~\ref{cartoon}.
We also thank N. Regnault, Z. Papic for useful comments and X. G. Wen for stimulating discussion.
This work is supported by the U.S. Department of Energy,
Office of Basic Energy Sciences under grants No. DE-FG02-06ER46305
(W.~Z. and D.~N.~S.) and DE-SC0002140 (F.~D.~M.~H. and Z.~L.), the latter also for the use of computational facilities at Princeton University.
Z.~L. was additionally supported by Alexander von Humboldt Research Fellowship for Postdoctoral Researchers.
F.~D.~M.~H. also acknowledges partial support from the W. M. Keck Foundation.


\bibliographystyle{apsrev}
\bibliography{bilayer_half}{}


\clearpage
\widetext
\appendix
\begin{appendices}

\section{Computational Methods}\label{app:method}


\subsection{Fractional Quantum Hall Bilayer Hamiltonian}
In a perpendicular external magnetic field, electrons moving in two spatial dimensions occupy highly-degenerate orbitals in each Landau level. When the magnetic field is strong, we can assume that electrons are spin-polarized (in experiments, the Zeeman energy is order of Kelvin, which is indeed much larger than the reported energy gap) and their dynamics is restricted to the orbitals in the lowest Landau level (LLL). Under these circumstances, the Hamiltonian of a fractional quantum Hall (FQH) bilayer system can be written as
\begin{eqnarray}\label{ham}
H&&=\sum_{\{m_i\}=0}^{N_{s}-1} \sum_{\sigma=\uparrow,\downarrow}V^{\sigma\sigma}_{m_1,m_2,m_3,m_4}
 c^\dagger_{m_1\sigma}c^\dagger_{m_2\sigma}c_{m_3\sigma}c_{m_4\sigma} \nonumber\\
&&+\sum_{\{m_i\}=0}^{N_{s}-1} \sum_{\sigma=\uparrow,\downarrow}V^{\sigma\bar{\sigma}}_{m_1,m_2,m_3,m_4}
 c^\dagger_{m_1\sigma}c^\dagger_{m_2\bar{\sigma}}c_{m_3\bar{\sigma}}c_{m_4\sigma}\nonumber\\
&&-t_{\perp}\sum_{m=0}^{N_s-1}\sum_{\sigma=\uparrow,\downarrow}c^\dagger_{m,\sigma}c_{m,\bar{\sigma}},
\end{eqnarray}
where $N_s$ is the total number of LLL orbitals in each layer, $c^{\dagger}_{m,\sigma} (c_{m,\sigma})$
is the creation (annihilation) operator of an electron in the LLL orbital $m$ of layer $\sigma (\bar{\sigma})=\uparrow(\downarrow),\downarrow(\uparrow)$, 
and $t_{\perp}$ describes the tunneling strength between two layers. $V^{\sigma\sigma}_{m_1,m_2,m_3,m_4}$ and $V^{\sigma\bar{\sigma}}_{m_1,m_2,m_3,m_4}$
are matrix elements of the intralayer and interlayer interaction, respectively, which can be computed by the standard second-quantization procedure once we adopt a specific geometry for the system. In the following, we give the details on the torus geometry and spherical geometry that we use in the main text.

\subsection{Torus Geometry}\label{app:method1}
The advantage of the torus geometry is its nonzero genus, which allows us to distinguish different topological orders by their
ground-state degeneracies.

We consider $N_e$ electrons moving on two rectangular tori with a perpendicular magnetic field. Each torus, corresponding to a layer, is spanned by
${\bf L}_1 = L_1 {\bf e}_x$ and ${\bf L}_2 = L_2 {\bf e}_y$, where ${\bf e}_x$ and ${\bf e}_y$ are fixed Cartesian unit vectors, and
$L_1$ and $L_2$ are lengths of the two fundamental cycles of the torus. Required by the magnetic translation invariance, the number of fluxes $N_\phi$ penetrating each torus, which is equal to the number of orbitals $N_s$ in one Landau level per layer, must be an integer $N_{s}=N_\phi=L_1L_2/(2\pi \ell^2)$. The total filling fraction in two layers is then defined as $\nu_T=N_e/N_\phi=N_e/N_s$.
In the following, we set the magnetic length $\ell=1$ as the length unit. 
In the Landau gauge ${\bf A}= Bx {\bf e}_y$, the basis of LLL single-particle states can be taken as
$\psi_j^\sigma(x_\sigma,y_\sigma)=\Big(\frac{1}{\sqrt{\pi}L_2}\Big)^{\frac{1}{2}}\sum_{n=-\infty}^{+\infty}
e^{\textrm{i}\frac{2\pi}{L_2}(j+nN_s)y_\sigma}e^{
-\frac{1}{2}[x_\sigma-\frac{2\pi}{L_2}(j+nN_s)]^2}$, where $(x_\sigma,y_\sigma)$ is the coordinate in layer $\sigma$ and
$j=0,1,\cdots,N_{s}-1$ is the orbital momentum.
Then the standard second-quantization procedures give
\begin{equation} \label{torusv}
V^{\sigma\sigma'}_{m_1,m_2,m_3,m_4}=\delta_{m_1+m_2,m_3+m_4}^{\textrm{mod} N_s}\frac{1}{4\pi N_s}\sum_{q_1,q_2=-\infty}^{+\infty}
\delta_{q_2,m_1-m_4}^{\textrm{mod} N_s}V_{\sigma\sigma'}(q_x,q_y)
e^{-\frac{1}{2}(q_x^2+q_y^2)}e^{\textrm{i}\frac{2\pi q_1}{N_s}(m_1-m_3)},
\end{equation}
where $q_x=\frac{2\pi q_1}{L_1}$, $q_y=\frac{2\pi q_2}{L_2}$, and
$V^{\sigma\sigma'}({\bf q})$ is the Fourier transform of the interaction in real space.

The detailed form of $V_{\sigma\sigma'}({\bf q})$ depends on the theoretical model of our bilayer FQH system.
In this work, we consider Coulomb-interacting electrons in double quantum wells, each of which is described by an infinite square well with width $w$
and separated from each other by distance $d$. Then we have
\begin{equation} \label{ham:intra}
 V_{\sigma\sigma}({\bf q})= \frac{1}{q} \frac{3qw+\frac{8\pi^2}{qw}-\frac{32\pi^4(1-e^{-qw})}{q^2w^2(q^2w^2+4\pi^2)}}{q^2w^2+4\pi^2}
\end{equation}
for the intralayer interaction, and
\begin{equation} \label{ham:inter}
V_{\sigma\bar{\sigma}}({\bf q})= \frac{1}{q} e^{-qd}
\end{equation}
for the interlayer interaction. These choices can well describe the experimental setups in GaAs/AlAs systems.

The magnetic translation invariance in two directions on the torus geometry allows us to label each many-body eigenstate of the Hamiltonian (\ref{ham})
by a two-dimensional momentum $(K_1,K_2)$.
In our exact diagonalization calculation, we utilize the full symmetry $(K_1,K_2)$. However, in the DMRG calculation,
we only use one quantum number $K_2$ (relabeled as $K$), which
is the total orbital momentum of the system.

\subsection{Spherical Geometry}
Compared with the torus geometry, spherical geometry has zero genus, thus we cannot distinguish different topological orders
by their ground-state degeneracies. However, the unique ground state and a single edge per layer for the orbital cut liberate us from the complicated ground-state
superposition and edge mode combination that may happen on the torus, thus
making the spherical geometry a particularly suitable platform for the entanglement spectroscopy.

We use Haldane's representation of the spherical geometry \cite{Haldane1983}. In our bilayer FQH system,
$N_e$ electrons are confined on the surfaces of two spheres. Each sphere, corresponding to a layer,
contains a magnetic monopole of strength $Q$.
The total number of magnetic fluxes through each spherical surface
is quantized to be an integer $N_\phi=2Q$.
The basis of LLL single-particle states can be taken as
$\psi_j^\sigma(u_\sigma,v_\sigma)=\sqrt{\frac{(2Q+1)!}{4\pi(Q+j)!(Q-j)!}}u_\sigma^{Q+j}v_\sigma^{Q-j}$
with orbital angular momentum $j=-Q,-Q+1,\cdots,Q$, thus there are $N_s=N_\phi+1=2Q+1$ orbitals in the LLL.
$(u_\sigma,v_\sigma)$ is the spinor variable in layer $\sigma$ with $u=\cos(\theta/2)e^{i\phi/2}$ and $v=\sin(\theta/2)e^{-i\phi/2}$,
where $\theta$ and $\phi$ are the spherical coordinates.
The total filling fraction in two layers is defined as $\nu_T=N_e/(N_\phi+\mathcal{S})=N_e/(N_s+\mathcal{S}-1)$,
where $\mathcal{S}$ is a finite-size shift on sphere.
Please note that both the $(331)$ Halperin state and Moore-Read Pfaffian state live in
$\mathcal{S}=3$. Standard second-quantization procedures lead to
\begin{eqnarray}
\label{eq:sphereV}
V^{\sigma\sigma'}_{m_1,m_2,m_3,m_4}=\delta_{m_1+m_2,m_3+m_4}\frac{1}{2}\sum_{l=0}^{2Q}\mathcal{V}_l^{\sigma\sigma'}[2(2Q-l)+1]
\left(
\begin{array}{ccc}
Q&Q&2Q-l\\
m_1-Q&m_2-Q&2Q-(m_1+m_2)
\end{array}\right)\nonumber\\
\times\left(
\begin{array}{ccc}
Q&Q&2Q-l\\
m_4-Q&m_3-Q&2Q-(m_3+m_4)
\end{array}\right),
\end{eqnarray}
where $m_{1,2,3,4}=0,1,\cdots,2Q$, $\left(
\begin{array}{ccc}
.&.&.\\
.&.&.
\end{array}\right)$ is the Wigner $3-j$ symbol, and $\mathcal{V}_l^{\sigma\sigma'}$ is the Haldane's pseudopotential parameter of the interaction.
For simplicity, we just use the LLL pseudopotential parameters on an infinite plane obtained by
\begin{equation}\label{eq:pseudo}
 \mathcal{V}_l^{\sigma\sigma'}=\frac{1}{(2\pi)^2}\int V_{\sigma\sigma'}({\bf q}) \mathcal{L}_l(q^2) e^{-q^2} d^2{\bf q},
\end{equation}
where $\mathcal{L}_l$ is the Laguerre polynomial, and $V_{\sigma\sigma'}({\bf q})$ is given by Eqs.~(\ref{ham:intra}) and (\ref{ham:inter}).

The symmetry that we use in our calculation is the conservation of the total orbital angular momentum $L_z$ on the sphere.

\subsection{Density-Matrix Renormalization Group}
In the main text, our calculations are based on the unbiased density-matrix renormalization group (DMRG) algorithm\cite{White,Shibata,Feiguin2008,JizeZhao2011}.
The technical details about DMRG in momentum space have been reported in our previous studies\cite{JizeZhao2011}.
There, it has been shown that, for the single-layer $\nu=1/3$ Laughlin state and the $\nu=5/2$ Moore-Read Pfaffian state, DMRG can get reliable results with very high accuracy in much larger systems than the limit of exact diagonalization.
Now, we find that DMRG also has excellent performance in our bilayer FQH system on the torus and spherical geometry.
We have obtained the ground state for the spherical (toroidal) system up to $N_e=24$ ($N_e=12$) electrons
by keeping up to $12000$ states, which leads to a truncation error smaller than $3\times 10^{-5}$ in the final sweep.
We also emphasize that, for the calculations on the torus geometry,
since we need to track two ground states in each momentum sector simultaneously,
the fully converged results are limited to $N_e=12$. Compared with the torus geometry,
the calculations on the spherical geometry can reach systems as large as $N_e=24$ within controlled accuracy.

\section{The Counting of Edge Excitations}\label{app:counting}
\subsection{Moore-Read Pfaffian State}
Here we analyze the degeneracy pattern of the edge excitation spectrum of the fermionic Moore-Read (MR) Pfaffian state.
Our analysis is based on root configurations\cite{Bernevig2008} on the sphere, which are also equivalent to configurations in the thin-torus limit\cite{Bergholtz2005,Seidel2005}. The degeneracy of the MR Pfaffian edge excitations is the same as the number of root configurations that satisfy a specific generalized exclusion rule, i.e., no more than $2$ fermions in $4$ consecutive orbitals.

In Tables~\ref{t1} and \ref{t2}, we count the root configurations that satisfy this rule.
We start from the initial root configuration without edge excitations, for example, $1100110011|0000$, which is just the root configuration of the MR Pfaffian state itself. ``$|$'' indicates the right edge, which is open so electrons can hop across to form edge excitations.
The root configurations with edge excitations must have larger angular momentum $L_z$ than the initial one.
We list all of them in terms of their relative angular momentum $\Delta L_z$ to the initial root configuration, for which $\Delta L_z=0$.
Note that root configurations and their counting are different for even number (Table~\ref{t1}) and odd number (Table~\ref{t2}) of electrons.

\begin{table*}[!h]
\caption{ In this table, we count the root configurations of the MR Pfaffian edge excitations with even $N_e$. The counting
is $1,1,3,5,\cdots$ in the $\Delta L_z=0,1,2,3,\cdots$ sector.}
\begin{ruledtabular}\label{t1}
\begin{tabular}{cccccccc}
$\Delta L_z=0$&$\Delta L_z=1$&$\Delta L_z=2$&$\Delta L_z=3$\\
\hline
$1100110011|0000$&$1100110010|1000$&$1100110010|0100$&$1100110010|0010$\\
                     &               &$1100110001|1000$&$1100110001|0100$\\
                     &               &$1100101010|1000$&$1100101010|0100$\\
                     &               &                  &$1100101001|1000$\\
                                     & &                &$1010101010|1000$\\
\end{tabular}
\end{ruledtabular}
\end{table*}

\begin{table*}[!h]
\caption{ In this table, we count the root configurations of the MR Pfaffian edge excitations with odd $N_e$.
The counting is $1,2,4,7,\cdots$ in the $\Delta L_z=0,1,2,3,\cdots$ sector.}
\begin{ruledtabular}\label{t2}
\begin{tabular}{cccccccc}
$\Delta L_z=0$&$\Delta L_z=1$&$\Delta L_z=2$&$\Delta L_z=3$\\
\hline
$110011001|0000$&$110011000|1000$&$110011000|0100$&$110011000|0010$\\
                &$110010101|0000$&$110010100|1000$&$110010100|0100$\\
                   &             &$110010011|0000$&$110010010|1000$\\
                   &             &$101010101|0000$&$110001100|1000$\\
                   & &                            &$101010100|1000$\\
                   & &                            &$101010011|0000$\\
                   & &                          &$1010101010101|0000$\\
\end{tabular}
\end{ruledtabular}
\end{table*}

The counting of edge excitations given above is saturated only in the thermodynamic limit. In finite systems, we can only observe part of them.
For example, in Table \ref{t2}, the root configuration $1010101010101|0000$ in the $\Delta L_z=3$ sector requires at least $7$ electrons in the system.
So this excitation cannot be observed in smaller system sizes.

We can also count the edge excitation modes from the effective edge Hamiltonian.
The edge excitation of the MR Pfaffian state contains one branch of free bosons and one branch of Majorana fermions (Also see Appendix Sec. \ref{app:theory2})
with either periodic or antiperiodic boundary conditions. For free bosons plus antiperiodic Majorana fermions
(which corresponds to the ground state on the sphere), the excitation
spectrum is described by the Hamiltonian \cite{Zhao2012}
\begin{equation}
H_{\textrm{edge}}^{\textrm{AP}}=\sum_{m>0}[E_b(m)b_m^\dagger
b_m+E_f(m-1/2)c_{m-1/2}^\dagger c_{m-1/2}],
\label{HAP}
\end{equation}
where $b$ and $b^\dagger$ ($c$ and $c^\dagger$) are standard boson
(fermion) creation and annihilation operators, $E_b(m)$ [$E_f(m)$]
is the dispersion relation of bosons (fermions) and the total
momentum operator is defined as $K=\sum_{m>0}[mb_m^\dagger
b_m+(m-1/2)c_{m-1/2}^\dagger c_{m-1/2}]$. The degeneracy of the
edge excitations is the same as the number of energy levels in each $K$ sector,
and depends on the parity of the number of fermions
$(-1)^F,F=\sum_{m>0}c_{m-1/2}^\dagger c_{m-1/2}$. For even $F$, the
counting is $1,1,3,5,10,\cdots$ at $\Delta K=0,1,2,3,4,\cdots$;
while for odd $F$, the counting is $1,2,4,7,13,\cdots$ at
$\Delta K=0,1,2,3,4,\cdots$. Here $\Delta K$ is defined as $K-K_0$
where $K_0$ is the lowest momentum ($K_0=0$ for even $F$ and
$K_0=1/2$ for odd $F$). One can see that this method reaches exactly the same counting as that obtained by root configurations.

\subsection{$(331)$ Halperin State}\label{app:counting2}
In bilayer FQH systems, it is convenient to consider the orbital $m$ in the upper layer and the orbital $m$ in the lower layer as a site with four possible occupations: $0$ (no electrons), $\uparrow$ (one electron in the upper layer), $\downarrow$ (one electron in the lower layer), and $2$ (two electrons, one in each layer). In this site basis, the root configuration of the $(331)$ state on the sphere is $XX00XX00\cdots XX00XX$ \cite{Seidel2008}, where $XX\equiv(\uparrow\downarrow+\downarrow\uparrow)/\sqrt{2}$ is the triplet
between two nearest neighbour sites. The root configurations of the $(331)$ state and its edge excitations obey the following generalized exclusion rule: (1) there is no more than one electron in three consecutive orbitals within each layer; (2) the configuration of $2$ is forbidden; (3) electrons on two nearest neighbour sites must form the triplet $XX$. Some configurations, for example $XX02,XX0XX,XX\uparrow,XX\downarrow,XX0\uparrow,XX0\downarrow,\uparrow\uparrow,\uparrow0\uparrow,\downarrow\downarrow,\downarrow0\downarrow$ violate this generalized exclusion rule, thus they cannot appear in the root configurations. With this generalized exclusion rule, we can count the root configurations of the $(331)$ edge excitations, as shown in Tables~\ref{t3} and \ref{t4}. Again, the root configurations and their counting are different for even number (Table~\ref{t3}) and odd number (Table~\ref{t4}) of electrons.

\begin{table*}[!h]
\caption{In this table, we count the root configurations of the $(331)$ edge excitations with even $N_e$. The counting is $1,2,7,14,\cdots$ in the $\Delta L_z=0,1,2,3,\cdots$ sector. We also give the pseudospin quantum number $S_z=(N_e^\uparrow-N_e^\downarrow)/2$ for each root configuration.}
\begin{ruledtabular}\label{t3}
\begin{tabular}{cccccccc}
$\Delta L_z=0$&$\Delta L_z=1$&$\Delta L_z=2$&$\Delta L_z=3$\\
\hline
$XX00XX00XX|000$ $(S_z=0)$& $XX00XX00\uparrow0|\downarrow00$ $(S_z=0)$ & $XX00XX00\uparrow0|0\downarrow0$ $(S_z=0)$ & $XX00XX00\uparrow0|00\downarrow$ $(S_z=0)$ \\
                          & $XX00XX00\downarrow0|\uparrow00$ $(S_z=0)$ & $XX00XX00\downarrow0|0\uparrow0$ $(S_z=0)$ & $XX00XX00\downarrow0|00\uparrow$ $(S_z=0)$ \\
                          &                                            & $XX00XX000X|X00$ $(S_z=0)$                 & $XX00XX000\uparrow|0\downarrow0$ $(S_z=0)$ \\
                          &                                            & $XX00\uparrow0\downarrow0\uparrow0|\downarrow00$ $(S_z=0)$ & $XX00XX000\downarrow|0\uparrow0$ $(S_z=0)$ \\
                          &                                            & $XX00\downarrow0\uparrow0\downarrow0|\uparrow00$ $(S_z=0)$ & $XX00\uparrow0\downarrow0\uparrow0|0\downarrow0$ $(S_z=0)$ \\
                          &                                            & $XX00XX00\uparrow0|0\uparrow0$ $(S_z=1)$                   & $XX00\downarrow0\uparrow0\downarrow0|0\uparrow0$ $(S_z=0)$  \\
                          &                                            & $XX00XX00\downarrow0|0\downarrow0$ $(S_z=-1)$              & $XX00\uparrow0\downarrow00X|X00$ $(S_z=0)$  \\
& & & $XX00\downarrow0\uparrow00X|X00$ $(S_z=0)$  \\
& & & $\uparrow0\downarrow0\uparrow0\downarrow0\uparrow0|\downarrow00$ $(S_z=0)$  \\
& & & $\downarrow0\uparrow0\downarrow0\uparrow0\downarrow0|\uparrow00$ $(S_z=0)$  \\
& & & $XX00XX00\uparrow0|00\uparrow$ $(S_z=1)$  \\
& & & $XX00\uparrow0\downarrow0\uparrow0|0\uparrow0$ $(S_z=1)$  \\
& & & $XX00XX00\downarrow0|00\downarrow$ $(S_z=-1)$  \\
& & & $XX00\downarrow0\uparrow0\downarrow0|0\downarrow0$ $(S_z=-1)$  \\

\end{tabular}
\end{ruledtabular}
\end{table*}

\begin{table*}[!h]
\caption{In this table, we count the root configurations of the $(331)$ edge excitations with odd $N_e$.
The counting is $2,4,10,\cdots$ in the $\Delta L_z=0,1,2,\cdots$ sector. We also give the pseudospin quantum number $S_z=(N_e^\uparrow-N_e^\downarrow)/2$ for each root configuration.}
\begin{ruledtabular}\label{t4}
\begin{tabular}{cccccccc}
$\Delta L_z=0$&$\Delta L_z=1$&$\Delta L_z=2$\\
\hline
$XX00XX00\uparrow|000$ $(S_z=1/2)$   &$XX00XX000|\uparrow00$ $(S_z=1/2)$                   &$XX00XX000|0\uparrow0$ $(S_z=1/2)$ \\
$XX00XX00\downarrow|000$ $(S_z=-1/2)$&$XX00\uparrow0\downarrow0\uparrow|000$ $(S_z=1/2)$   &$XX00\uparrow0\downarrow00|\uparrow00$ $(S_z=1/2)$ \\
                                     &$XX00XX000|\downarrow00$ $(S_z=-1/2)$                &$XX00\downarrow0\uparrow00|\uparrow00$ $(S_z=1/2)$ \\
                                     &$XX00\downarrow0\uparrow0\downarrow|000$ $(S_z=-1/2)$&$XX00\uparrow00XX|000$ $(S_z=1/2)$ \\
                                     &                                                     &$\uparrow0\downarrow0\uparrow0\downarrow0\uparrow|000$ $(S_z=1/2)$ \\
                                     &                                                     &$XX00XX000|0\downarrow0$ $(S_z=-1/2)$ \\
                                     &                                                     &$XX00\uparrow0\downarrow00|\downarrow00$ $(S_z=-1/2)$  \\
                                     &                                                     &$XX00\downarrow0\uparrow00|\downarrow00$ $(S_z=-1/2)$  \\
                                     &                                                     &$XX00\downarrow00XX|000$ $(S_z=-1/2)$  \\
                                     &                                                     &$\downarrow0\uparrow0\downarrow0\uparrow0\downarrow|000$ $(S_z=-1/2)$  \\
\end{tabular}
\end{ruledtabular}
\end{table*}

\section{Additional Results of Entanglement Spectra} \label{app:ES}
\begin{figure}[!htb]
 \begin{minipage}{0.98\linewidth}
 \centering
 \includegraphics[width=5.0in]{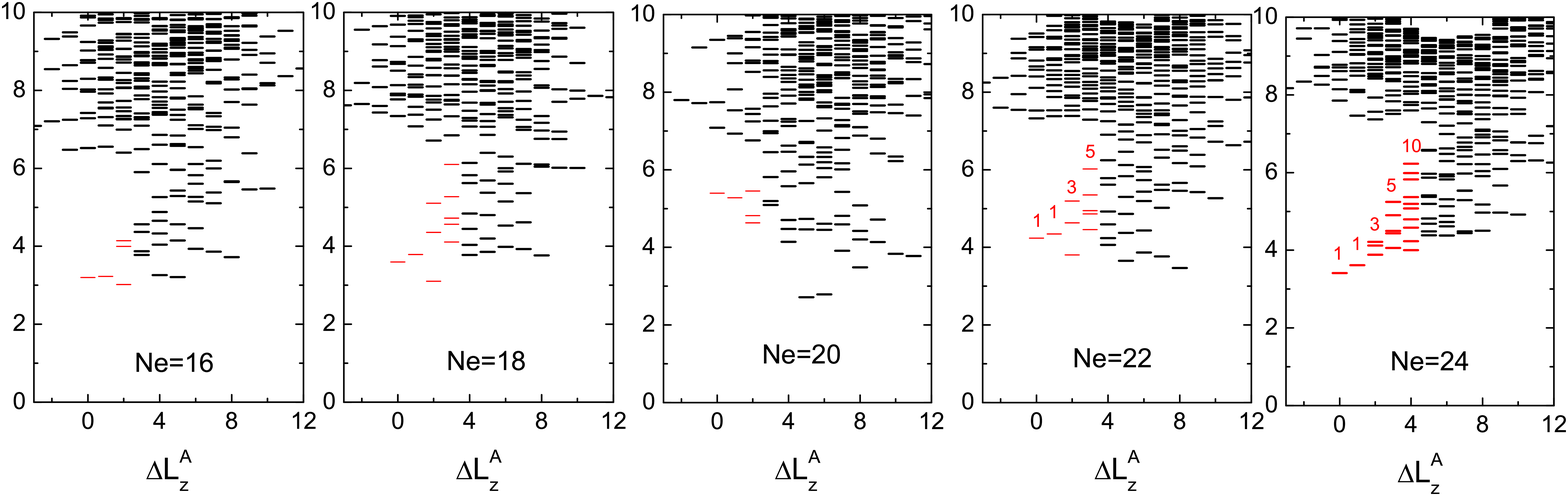}
 \includegraphics[width=5.0in]{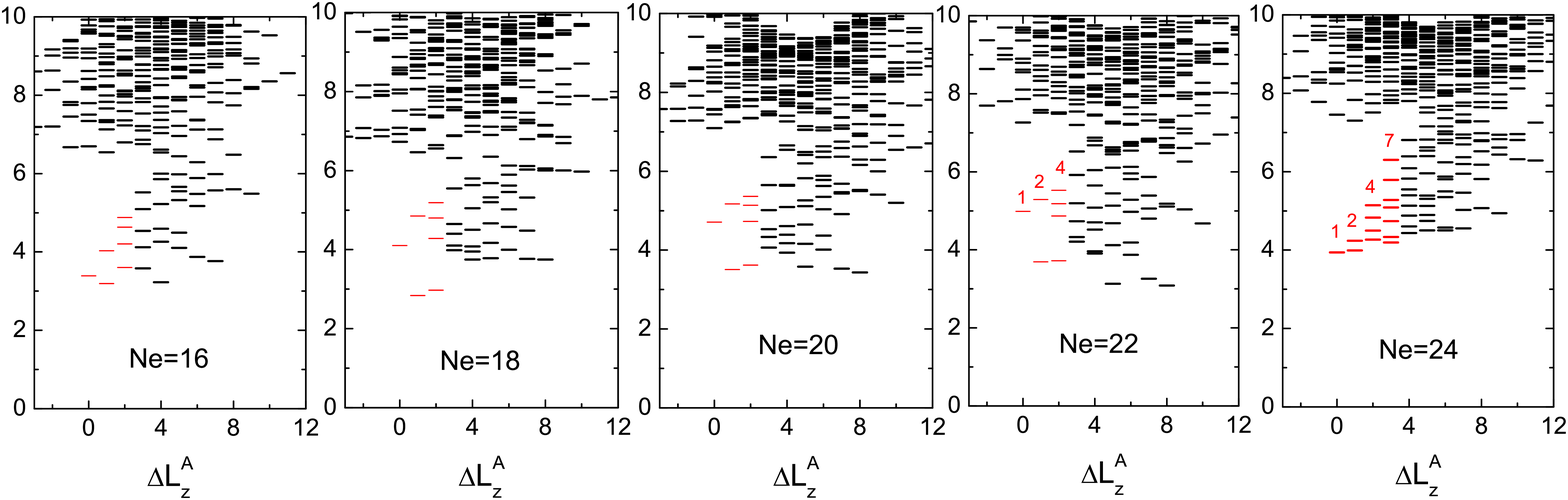}
 \end{minipage}
 \caption{The low-lying orbital ES for various bilayer system sizes $N_e=16,18,20,22,24$ at the tunneling strength $t_\perp=0.10$, layer width $w=1.5$ and layer distance $d=3.0$,
 with even (top) or odd (bottom) number of electrons in the half-cut subsystem.
 The levels whose counting is consistent with the MR Pfaffian edge excitations proposed in Tables~\ref{t1} and \ref{t2} are labeled by red.
 $\Delta L^A_z=L^A_z-L^A_{z,\min}$, where $L^A_{z,\min}$ is the total angular momentum of the subsystem $A$ without edge excitations.} \label{sfig:ES_MR_Ne}
\end{figure}

An artificial edge is produced by the orbital cut of the whole system into two parts. The low-lying entanglement spectrum (ES) mimics the edge excitation spectrum of one subsystem across the cutting  edge. Thus the counting structure in the ES can be predicted by applying the analysis in Appendix Sec.~\ref{app:counting} to that subsystem, whose initial root configuration is the corresponding subsystem part of the root configuration of the whole system. Here we show the ground-state orbital ES in the MR Pfaffian phase of our bilayer FQH system for various system sizes $N_e=16,18,20,22$ and $24$ (Fig.~\ref{sfig:ES_MR_Ne}).
One can see that the leading ES counting at all system sizes always displays $1,1,3$ ($1,2,4$) for even (odd) number of electrons in the half-cut subsystem, matching the predictions in Tables~\ref{t1} and \ref{t2}.
For $N_e=20$, we observe a relatively small entanglement gap,
which might be attributed to that this system size is ``aliased'' to another possible FQH state at $\nu=4/7$ on the sphere.
If we go to larger systems like $N_e=22$ and $24$, the entanglement gap becomes stronger again.
Here, the low-lying ES structure is robust against finite-size effect,
providing a fingerprint of the non-Abelian MR Pfaffian nature of the ground state.

\begin{figure}[!htb]
 \begin{minipage}{0.98\linewidth}
 \centering
 \includegraphics[width=0.98\linewidth]{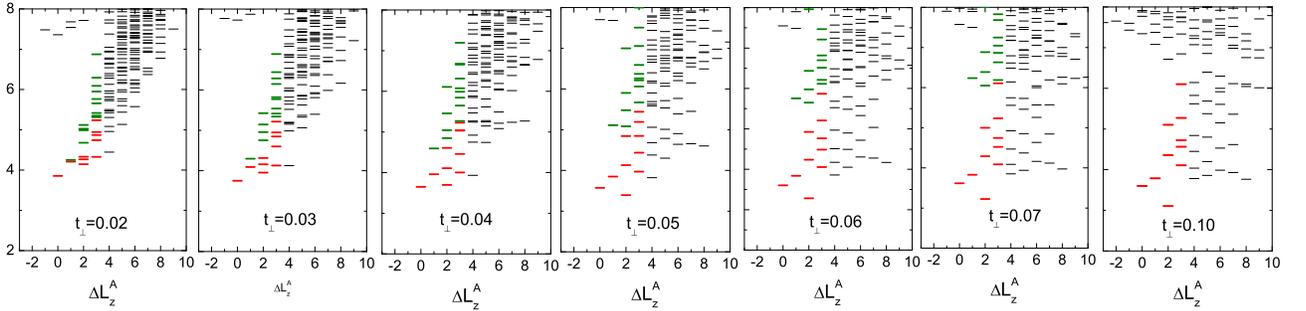}
 \end{minipage}
 \caption{The low-lying orbital ES of $N_e=18$ for different tunneling strength $t_\perp$ at layer width $w=1.5$ and layer distance $d=3.0$,
 with even number of electrons in the half-cut subsystem.
 The levels whose counting is consistent with the MR Pfaffian edge excitations proposed in Tables~\ref{t1} and \ref{t2} are labeled by red.
 The green levels match the $(331)$ edge excitations proposed in Tables~\ref{t3} and \ref{t4} at small $t_\perp$, but are continuously gapped out
 with increasing the tunneling strength.
 $\Delta L^A_z=L^A_z-L^A_{z,\min}$, where $L^A_{z,\min}$ is the total angular momentum of the subsystem $A$ without edge excitations.} \label{sfig:ES_MR_t}
\end{figure}

We also track the evolution of the ground-state orbital ES as a function of the tunneling strength $t_\perp$.
In Fig.~\ref{sfig:ES_MR_t}, we show the ES by varying $t_\perp$ from $0.02$ to $0.10$ at layer width $w=1.5$ and layer distance $d=3.0$.
In the weak-tunneling regime ($t_\perp<0.04$),
the leading ES counting matches the expectation of the $(331)$ Halperin state in Tables~\ref{t3} and \ref{t4}.
Remarkably, with increasing $t_\perp$, some levels in angular momentum sectors
$\Delta L_z^A \geq 1$ are being continuously gapped out.
For example, at $t_\perp=0.05$, the gap between the lowest level and the second lowest level in the $\Delta L_z^A=1$ sector becomes visible,
indicating the MR Pfaffian ES is developing.
The fact that some edge modes in the ES of the $(331)$ Halperin state are continuously being gapped out with increasing the tunneling strength $t_\perp$ is
consistent with the effective edge theory described in Appendix Sec.~\ref{app:theory2}.

\newpage
\section{Theoretical Consideration}\label{app:theory}
In this section, we review several different theories to understand the relation between the non-Abelian MR Pfaffian state
and the Abelian $(331)$ Halperin state. First, with the help of Cauchy identity, we show that the antisymmetrized
$(331)$ Halperin wave function leads to the MR Pfaffian wavefunction. Second, by including the tunneling effect,
it is plausible to reach the edge theory of the MR Pfaffian state from that of the $(331)$ Halperin state via gapping out
one branch of Majorana fermion from the neutral mode.
Third, working in the thin-torus limit, the quantum phase transition from the $(331)$ Halperin state to the
MR Pfaffian state can be captured by an effective one-dimensional transverse-field Ising model,
which helps us to elucidate the nature of the transition.  

\subsection{Model Wave Function}
It has been a long time since the discovery of the exact equivalence\cite{Greiter1992,Ho1995} between the MR Pfaffian wave function\cite{Moore}
and the antisymmetrized $(331)$ Halperin wave function \cite{halperin1983}.
Specifically, the MR Pfaffian wave function can be written as (we discard the Gaussian exponential factor hereafter)
\begin{align} \label{eq:Psi_MR}
\Psi_{\rm MR} = \prod_{i<j} (z_i - z_j)^2 {\rm Pf}\left(\frac{1}{z_{i} - z_{j}}\right) ,
\end{align}
where $z_i$'s are two-dimensional coordinates of electrons, and ${\rm Pf}(M_{ij})={\cal A}(M_{12}M_{34}\cdots M_{N-1,N})$ with ${\cal A}$ the antisymmetrization operator.
The $(331)$ Halperin wave function is
\begin{align} \label{eq:Psi_331}
\Psi_{331} = \prod_{i<j}  (z_i^{\uparrow} - z_j^{\uparrow})^3
\prod_{k<l} (w_k^{\downarrow} - w_l^{\downarrow})^3
\prod_{m,n}  (z_m^{\uparrow} - w_n^{\downarrow}) ,
\end{align}
where $z_i^\uparrow$'s and $w_i^{\downarrow}$'s are coordinates of electrons in the top and bottom layers denoted by the pseudospin indices $\uparrow$ and $\downarrow$, respectively.
It is important to note that the $(331)$ Halperin wavefunction can be analytically cast in to a paired form
\begin{equation*}
\Psi_{331} = \prod_{i<j} (x_i-x_j)^2 \det \Big[ \frac{1}{ z_i^{\uparrow}-w_j^{\downarrow} } \Big]
\end{equation*}
with the help of the Cauchy identity\cite{Greiter1992}
$\frac{\prod_{i<j}  (z_i^{\uparrow} - z_j^{\uparrow}) \prod_{k<l} (w_k^{\downarrow} - w_l^{\downarrow})}{\prod_{m,n}  (z_m^{\uparrow} - w_n^{\downarrow}) }=
\det \Big[ \frac{1}{ z_i^{\uparrow}-w_j^{\downarrow} } \Big]$,
where $\{x_i\}$ includes all $z_i^{\uparrow}$'s and $w_i^{\downarrow}$'s. Then, a further antisymmetrization precisely produces the MR Pfaffian wavefunction (up to a constant normalization factor), leading to
\begin{align}
\label{eq:MR_A331_equivalence}
\Psi_{\rm MR} = {\cal A} \Psi_{331}.
\end{align}


\subsection{Effective Edge Theory}\label{app:theory2}
Another efficient way to investigate the possible transition between the $(331)$ Halperin state and the MR Pfaffian state is
the effective edge theory. The key idea is that, starting from
the $(331)$ edge theory described by two chiral boson fields [central charge $c=2$ in conformal field theory (CFT)],
the tunneling effect between two layers tends to replace one boson field by a Majorana fermion field carrying $c=1/2$, while the other $c=1$ boson field is remained.

More precisely, we start from the edge theory of the $(331)$ Halperin state.
The gapless excitations are confined to two edges of the droplet, described by the action\cite{Wen1992}
\begin{equation*}
  S_{\textrm{edge}}=\frac{1}{4\pi} \int dt dx [K_{IJ}\partial_t u_I \partial_x u_J - V_{IJ}\partial_x u_I \partial_x u_J]
\end{equation*}
and the Hamiltonian
\begin{equation}\label{edge:ham}
  H_{\textrm{edge}}=\frac{1}{4\pi} \int dt dx V_{IJ} \partial_x u_I \partial_x u_J.
\end{equation}
The matrix $K_{IJ}$ which characterizes the topological properties of the $(331)$ Halperin state has the form of
\begin{equation}\label{edge:Kmatrix}
  K=\left(
      \begin{array}{cc}
        3 & 1 \\
        1 & 3 \\
      \end{array}
    \right).
\end{equation}
$u_I(t,x)$ ($I=1,2$ corresponding to two layers) are chiral bosonic fields describing two edge currents along the $x-$direction, which
satisfy the equal-time commutation relation
\begin{equation}\label{edge:communtation}
  [u_I(t,x),u_J(t,x')]=i\pi K_{IJ} {\rm sgn}(x-x').
\end{equation}
The matrix $V_{IJ}$ which contains the information of interactions between the edges has the form of
\begin{equation}\label{}
  V=\left(
      \begin{array}{cc}
        v & g \\
        g & v \\
      \end{array}
    \right),
\end{equation}
where we require $g^2<v^2$ so $V$ is positive definite.

$H_{\textrm{edge}}$ in Eq.~(\ref{edge:ham}) can be simplified by an orthogonal transformation
on the chiral bosonic fields
\begin{equation}\label{edge:orthotrans}
  \left(
    \begin{array}{c}
      u_1 \\
      u_2 \\
    \end{array}
  \right) =
  \left(
    \begin{array}{cc}
      \sqrt{2} & -1 \\
      \sqrt{2} & 1 \\
    \end{array}
  \right)
  \left(
  \begin{array}{c}
    \phi_c \\
    \phi_n
  \end{array}
  \right),
\end{equation}
leading to
\begin{equation}
  H_{\textrm{edge}}=\frac{1}{4\pi} \int dt dx [ v_c (\partial_x \phi_c)^2 + v_n (\partial_x \phi_n )^2 ].
\end{equation}
We refer to the new bosonic fields $\phi_c$ and $\phi_n$ as the charged and neutral edge mode, respectively \cite{Naud2000,Fradkin_Book}.
$\phi_c$ is related to the total electric charge on the two edges with velocity $v_c=4(v+g)$,
while $\phi_n$ is related to
the difference between two edges with velocity $v_n=2(v-g)$. In terms of the new fields, the commutators are now
independent:
\begin{equation}\label{edge:communtation}
  [\phi_c(t,x),\phi_c(t,x')]=i\pi {\rm sgn}(x-x'),\,\, [\phi_n(t,x),\phi_n(t,x')]=i\pi {\rm sgn}(x-x'),\,\, [\phi_c(t,x),\phi_n(t,x')]=0.
\end{equation}

Next we assume that electrons can tunnel between two edges, and the tunneling Hamiltonian takes the form of
\begin{equation}\label{edge:tunnel}
  H_{\textrm{tunnel}}=-t_{\perp} \int dt dx [\hat \psi^{\dagger}_1 \hat \psi_2+{\rm h.c.}],
\end{equation}
where $\hat \psi_{I=1,2}$ are the electron operators and satisfy the usual fermionic anti-commutation relation.
The relationship between the electron operators $\hat \psi_I$ and the chiral boson fields $u_I(t,x)$ is\cite{Wen1990c}
\begin{eqnarray}\label{}
  \hat \psi^{\dagger}_1= \eta e^{i u_1(t,x)} = \eta e^{i (\sqrt{2}\phi_c-\phi_n)}, \nonumber\\
  \hat \psi^{\dagger}_2= \eta e^{i u_2(t,x)} = \eta e^{i (\sqrt{2}\phi_c+\phi_n)},
\end{eqnarray}
where we have used Eq.~(\ref{edge:orthotrans}) and $\eta$ is a constant depending on the cutoff.
Then, the total Hamiltonian can be expressed as
\begin{eqnarray*}
  H &=& H_{\textrm{edge}} + H_{\textrm{tunnel}} = H_c+ H_n, \\
  H_c &=& \frac{1}{4\pi} \int dt dx v_c (\partial_x \phi_c)^2, \\
  H_n &=& \frac{1}{4\pi} \int dt dx [v_n (\partial_x \phi_n)^2 - t_{\perp}' (e^{2i\phi_n} + {\rm h.c.})],
\end{eqnarray*}
where $t_{\perp}'$ is a constant proportional to $t_{\perp}$. One can see that the tunneling only appears in the neutral mode Hamiltonian $H_n$.

The next important step is to fermionize $H_n$ by a Dirac fermion field $\psi_D\equiv\frac{1}{\sqrt{2\pi}}e^{i\phi_n}$, which
can be further decomposed in terms of two chiral Majorana fermion fields $\chi_{i=1,2}$ \cite{Naud2000} by $\psi_D=\frac{1}{\sqrt{2}} (\chi_1+i\chi_2)$.
Finally, we have
\begin{equation}\label{}
  H_n=-\frac{i}{2}\int dt dx [(v_n- t_{\perp}'') \chi_1\partial_x \chi_1 +
                               (v_n+t_{\perp}'') \chi_2\partial_x \chi_2],
\end{equation}
where $t_{\perp}''$ is a rescaled tunneling strength. Now two majorana fields are decoupled with different velocities modified by the tunneling.
The key observation is, when the condition $v_n- t_{\perp}''=0$ is satisfied,
one Majorana field $\chi_1$ vanishes, leading to
\begin{equation}\label{}
  H=\int dt dx [-\frac{i}{2}(v_n+ t_{\perp}'') \chi_2\partial_x \chi_2 +\frac{1}{4\pi}
                               v_c (\partial_x \phi_c)^2].
\end{equation}
Physically, it means one majorana mode
can be completely gapped out with the help of the tunneling effect.
The remaining edge theory includes a chiral boson (charged mode, $\phi_c$) with central charge $c_\phi=1$, and
a chiral Majorana fermion (neutral mode, $\chi_1$) with central charge $c_\chi=1/2$. Thus, the total central charge of remaining system is
$c_{{\rm eff}}=c_\phi+c_\chi=1+1/2$,
which is consistent with the expectation of the MR Pfaffian state.


\subsection{Effective Theory in the Thin-Torus Limit}
In this section, we study the quantum phase transition from the $(331)$ Halperin state to the MR Pfaffian state
driven by the interlayer tunneling from a different perspective.
That is, we will derive an effective theory for the underlying quantum phase
transition in the thin-torus limit\cite{Bergholtz2005,Seidel2005,Seidel2008,Bernevig2008}.
Unlike the effective edge theory,
such kind of effective theory is constructed for the bulk and is expected to describe how the ground-state manifold evolves from the $(331)$ degeneracy to the MR Pfaffian degeneracy.

Recall that the interaction matrix elements $V^{\sigma\sigma'}_{m_1,m_2,m_3,m_4}$ on the torus only depends on $m_1-m_3$ and $m_1-m_4$ [Eq.~(\ref{torusv})]. This allows us to reformulate the translation invariant interaction Hamiltonian in Eq.~(\ref{ham}) as
\begin{eqnarray}
H_{{\rm int}}&=&\sum_{\sigma,\sigma'=\uparrow,\downarrow}\sum_{i=0}^{N_s-1}\sum_{r,s} U^{\sigma\sigma'}_{r,s}
 c^\dagger_{i+s,\sigma}c^\dagger_{i+r,\sigma'}c_{i+s+r,\sigma'}c_{i,\sigma}.
\end{eqnarray}
Since the single-particle LLL wave function on the torus
$\psi_j^\sigma(x_\sigma,y_\sigma)=\Big(\frac{1}{\sqrt{\pi}L_2}\Big)^{\frac{1}{2}}\sum_{n=-\infty}^{+\infty}
e^{\textrm{i}\frac{2\pi}{L_2}(j+nN_s)y_\sigma}e^{
-\frac{1}{2}[x_\sigma-\frac{2\pi}{L_2}(j+nN_s)]^2}$
is localized along $x_\sigma=2\pi j/L_2$, the separation of two consecutive orbitals,
or the overlap between $\psi_j$ and $\psi_{j+1}$, is controlled by a single parameter $\kappa = 2\pi/L_2$.
In the thin-torus limit $L_2 \ll1$ or $\kappa \gg 1$, the overlap between two adjacent Landau orbitals is negligible, thus the system
can be viewed as a one-dimensional chain.
Because the magnitude of $U^{\sigma\sigma'}_{r,s}\propto e^{-\kappa^2(s^2+r^2)/2}$ decays exponentially when $\kappa\rightarrow\infty$,
the dominated interaction Hamiltonian in the thin-torus limit is\cite{Bergholtz2005,Seidel2005} 
\begin{eqnarray}\label{tth}
H_{\rm int}=\sum_{\sigma,\sigma'}\sum _{i} \sum'_{r} U^{\sigma\sigma'}_{r,0} n^{\sigma}_{i}  n^{\sigma'}_{i+r},
\end{eqnarray}
where $n^{\sigma}_{i}=c^\dagger_{i,\sigma}c_{i,\sigma}$, and $\sum'_{r}$ means $r=0$ is excluded if $\sigma=\sigma'$. $U^{\sigma\sigma'}_{r,0}$ can be treated perturbatively with the increase of $r$.

The thin-torus interaction (\ref{tth}) only includes exponentially decaying electrostatic terms. Once we truncate it at short ranges, the ground states at a fixed filling fraction have simple charge-density-wave patterns, i.e., the thin-torus configurations of the corresponding FQH states. Here we give the solutions for the $(331)$ Halperin state (Table~\ref{t5}) and the MR Pfaffian state (Table~\ref{t6}) [For the MR Pfaffian state, we need a similar thin-torus analysis of its three-body parent Hamiltonian rather than (\ref{tth})], labeled by their momentum quantum numbers $(K_1,K_2)$ (see Appendix Sec.~\ref{app:method1}) on the torus. Actually they are quite similar to each other, except that the $(331)$ Halperin state takes layer indices and has one more configuration $0\uparrow0\downarrow\cdots0\uparrow0\downarrow-0\downarrow0\uparrow\cdots0\downarrow0\uparrow$ in the $(K_1,K_2)=(0,0)$ sector. As we will analyze in the next section Appendix Sec.~\ref{app:ED2}, strong tunneling favors the symmetric basis $c_{m,{\rm s}}^\dagger=\frac{1}{\sqrt{2}}(c_{m\uparrow}^\dagger+c_{m\downarrow}^\dagger)$. Therefore, $0\uparrow0\downarrow\cdots0\uparrow0\downarrow-0\downarrow0\uparrow\cdots0\downarrow0\uparrow$ vanishes at strong tunneling, because both $0\uparrow0\downarrow\cdots0\uparrow0\downarrow$ and $0\downarrow0\uparrow\cdots0\downarrow0\uparrow$ are mapped to $0101\cdots0101$ in the effective single-component orbitals under the symmetric basis. Other three thin-torus configurations of the $(331)$ Halperin state are mapped to $0101\cdots0101$, $1100\cdots1100+0011\cdots0011$ and $1100\cdots1100-0011\cdots0011$, respectively, which exactly matches the thin-torus configurations of the MR Pfaffian state. This means, at least in the thin-torus limit, interlayer tunneling is indeed a mechanism that can induce the quantum phase transition from the $(331)$ Halperin state to the MR Pfaffian state.

\begin{table*}[!h]
\caption{The thin-torus configurations of the $(331)$ Halperin state, expressed in the bilayer FQH site basis (see Appendix Sec.~\ref{app:counting2}). Here we neglect the center-of-mass degeneracy caused by the translation of each configuration.}
\begin{ruledtabular}\label{t5}
\begin{tabular}{cccc}
 & Thin-torus configuration & $(K_1,K_2)$ \\
\hline
& $0 \uparrow 0 \downarrow \cdots0 \uparrow 0 \downarrow -0 \downarrow 0\uparrow\cdots 0 \downarrow 0 \uparrow $  & $(0,0)$ \\
& $0 \uparrow 0 \downarrow\cdots 0 \uparrow 0 \downarrow +0 \downarrow 0\uparrow\cdots 0 \downarrow 0 \uparrow $  & $(\pi,0)$ \\
& $XX00\cdots XX00+00XX\cdots00XX$   &  $(0,\pi)$\\
& $XX00\cdots XX00-00XX\cdots00XX$   &  $(\pi,\pi)$\\
\end{tabular}
\end{ruledtabular}
\end{table*}

\begin{table*}[!h]
\caption{The thin-torus configurations of the MR Pfaffian state, expressed in the single-component FQH orbital basis. Here we neglect the center-of-mass degeneracy caused by the translation of each configuration.}
\begin{ruledtabular}\label{t6}
\begin{tabular}{cccc}
 & Thin-torus configuration &  $(K_1,K_2)$ \\
\hline
& $0 1 0 1\cdots 0 1 0 1 $  & $(\pi,0)$ \\
& $1 1 0 0\cdots 1 1 0 0 + 0011\cdots0011$   &  $(0,\pi)$\\
& $1 1 0 0\cdots 1 1 0 0 - 0011\cdots0011$   &  $(\pi,\pi)$\\
\end{tabular}
\end{ruledtabular}
\end{table*}

Now we start to build an effective bulk theory for the phase transition. We truncate the interaction Hamiltonian (\ref{tth}) at $r=2$, and add the tunneling term, leading to the total Hamiltonian as
\begin{eqnarray}
&&H= H_0 +H_1 + H_t, \nonumber\\
&&H_0= \sum_{i,\sigma}U^{\sigma\sigma}_{1,0} n^{\sigma}_{i}n^{\sigma}_{i+1}  +
       \sum_{i,\sigma}U^{\sigma\bar{\sigma}}_{0,0}n^{\sigma}_{i}n^{\bar{\sigma}}_{i}, \nonumber \\
&&H_1= \sum_{i,\sigma} U^{\sigma\sigma}_{2,0}n^{\sigma}_{i}n^{\sigma}_{i+2}+\sum_{i,\sigma} U^{\sigma\bar{\sigma}}_{1,0}n^{\sigma}_{i}n^{\bar{\sigma}}_{i+1}+\sum_{i,\sigma} U^{\sigma\bar{\sigma}}_{2,0}n^{\sigma}_{i}n^{\bar{\sigma}}_{i+2},  \nonumber  \\
&&H_t= -t_{\perp}\sum_{i} c_{i,\uparrow}^\dagger c_{i,\downarrow}+{\rm h.c.}.
\end{eqnarray}
We take $H_1+H_t$ as perturbation and construct an effective Hamiltonian in the degenerate ground-state manifold of $H_0$.

\begin{figure}[!htb]
 \begin{minipage}{0.98\linewidth}
 \centering
 \includegraphics[width=0.35\textwidth]{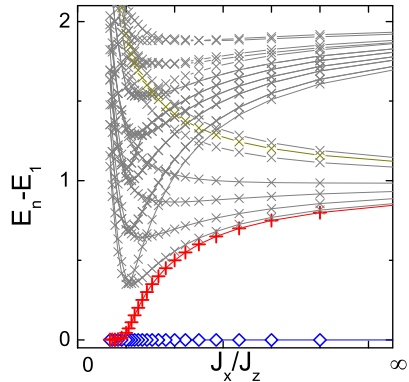}
 \end{minipage}
 \caption{Energy spectra of the one-dimensional transverse field Ising model as a function of parameter $J_x/J_z$.
 Here, the calculation is performed on a spin chain with $18$ sites. It is clear that, as increasing $J_x$, 
 one ground state is continuously gapped out (labeled by red cross).
 Please note that the evolution of the low-energy spectrum versus $J_x$ is very similar to that in our bilayer FQH system [Fig.~\ref{torus}(a)].
 Please see Ref.~\onlinecite{Sachdev2000} for detailed analysis of the transition in Ising model.
 } \label{sfig:Ising}
\end{figure}

Note that all configurations that can be related to $0 \uparrow 0 \downarrow\cdots 0 \uparrow 0 \downarrow$ by spin flips belong to the degenerate ground-state manifold of $H_0$. For these configurations, if we introduce a new basis $|+ \rangle_i \equiv  [0 \uparrow]_i  ,|- \rangle_i \equiv  [0 \downarrow]_i $ for a unit cell of two consecutive orbitals, and define $\sigma^x_i=|+ \rangle _{i}\langle -| + |- \rangle_{i} \langle +| ,
\sigma^z_i=|+ \rangle_{i} \langle +| - |- \rangle _{i}\langle -|$,
we can reach an effective Hamiltonian
\begin{eqnarray}\label{eq:Ising-1}
H_{\rm eff}=-J_x\sum_{i}\sigma_{i}^{x} + J_z \sum_{i} \sigma^z_{i}\sigma^z_{i+1}
\end{eqnarray}
up to the first-order perturbation, with effective coupling $J_x \sim t_\perp $ and $J_z\sim (U^{\sigma,\sigma}_{2,0}-U^{\sigma,\bar{\sigma}}_{2,0})/2$.
$H_{\rm eff}$ is nothing but the widely studied one-dimensional transverse field Ising model, which hosts two gapped phases \cite{Sachdev2000}.
The ground states are doubly degenerate for $J_x<J_z$, while there is only a unique $Z_2$ symmetric ground state for $J_x>J_z$. The transition at $J_x=J_z$ between these two phases has been determined to be continuous.
In the $J_x/J_z \rightarrow 0$ limit, the two ground states are $|+\rangle|-\rangle\cdots|+\rangle|-\rangle=0 \uparrow 0 \downarrow \cdots0 \uparrow 0 \downarrow$ and $|-\rangle|+\rangle\cdots|-\rangle|+\rangle=0 \downarrow 0\uparrow\cdots 0 \downarrow 0 \uparrow$, which
exactly match the $(331)$ thin-torus configurations with $(K_1,K_2)=(0,0)$ and $(\pi,0)$ up to a superposition.
In the $J_x/J_z \rightarrow \infty$ limit, the unique ground state is polarized in the $x-$direction with the form of $\prod_i \frac{|+\rangle_i+|-\rangle_i}{\sqrt{2}}=0 \rightarrow 0 \rightarrow\cdots 0 \rightarrow 0 \rightarrow$ with $\rightarrow\equiv\uparrow+\downarrow$, which exactly matches the MR Pfaffian thin-torus configuration with $(K_1,K_2)=(\pi,0)$. This effective model indicates that the $(331)$ configuration with $(K_1,K_2)=(0,0)$ can indeed be gapped out through a continuous phase transition by increasing the tunneling $t_{\perp}$ (Fig. \ref{sfig:Ising}).

In addition, all configurations that are related to $\uparrow\downarrow00\cdots\uparrow\downarrow00$ by simultaneously flipping two nearest neighbor spins also belong to the degenerate ground-state manifold of $H_0$. For these configurations, we can introduce a new basis $|+ \rangle_i \equiv  [\uparrow\downarrow00]_i  ,|- \rangle_i \equiv  [\downarrow\uparrow00]_i $ for a unit cell of four consecutive orbitals, and define $\sigma^x_i=|+ \rangle _{i}\langle -| + |- \rangle_{i} \langle +| ,
\sigma^z_i=|+ \rangle_{i} \langle +| - |- \rangle _{i}\langle -|$. The second-order perturbation leads to an effective Hamiltonian $H_{\rm eff}=-J_x\sum_i\sigma_i^x$ with $J_x\sim2t_\perp^2/U^{\sigma\sigma}_{1,0}$. Therefore, there is no phase transition and the ground state is always $\prod_i \frac{|+\rangle_i+|-\rangle_i}{\sqrt{2}}=XX00\cdots XX00$. The same conclusion also holds for those configurations that are related to $00\uparrow\downarrow\cdots00\uparrow\downarrow$ by simultaneously flipping two nearest neighbor spins. Therefore, the $(331)$ degeneracy in the $(0,\pi)$ and $(\pi,\pi)$ sectors cannot be changed by the tunneling.

We have clarified that the $(331)$ Halperin state and the MR Pfaffian state is
separated by a quantum critical point and the ground-state degeneracy
can be reduced from four-fold to three-fold.
To further elucidate the ``topological'' property of this phase transition,
we refer to the fermionic representation of the transverse field Ising model.
After the Wigner-Jordan transformation, one-dimensional transverse field Ising model
can be mapped to Kitaev chain model.
Thus, we reach an intriguing connection here: the quantum
phase transition from the $(331)$ Halperin state to the MR Pfaffian can be understood as a transition
from the weak $p-$wave pairing regime to the strong $p-$wave pairing regime.
The same statement was predicted years ago \cite{Read2000},
where Read and Green derived it by setting up the BCS effective quasiparticle Hamiltonian and
Bogoliubov transformation.
Here, we reach the same conclusion through the perturbation theory in the thin-torus limit.
In our approach,   the nature of transition becomes transparent through mapping the bilayer system to an exactly solvable model.


\section{Energy Spectra from Exact Diagonalization}\label{app:ED}
\subsection{Bilayer System}
In the main text, we show the energy spectra of $N_e=12$ on the torus obtained by DMRG.
Here we would like to present the torus energy spectra of smaller systems that can be reached by exact diagonalization (ED).
In our bilayer FQH system, the Hilbert space grows very fast with the increase of the system size, so ED calculations are strongly limited.

\begin{figure}[!htb]
 \begin{minipage}{0.98\linewidth}
 \centering
 \includegraphics[width=0.8\textwidth]{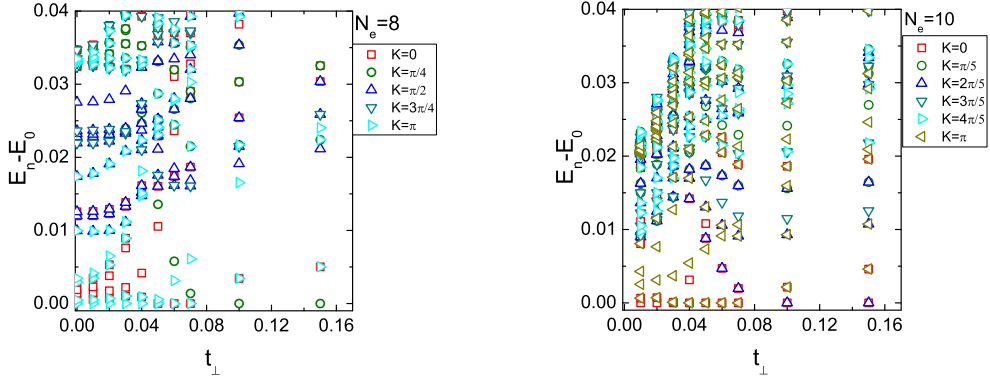}
 \end{minipage}
 \caption{ Energy spectra as a function of tunneling $t_{\perp}$ on a square torus with $N_e=8$ (left) and $10$ (right) electrons obtained by exact diagonalization.
 Different momentum sectors are labeled by different symbols. All calculations are performed at layer width $w=1.5$ and layer distance $d=3.0$.
 } \label{sfig:torus}
\end{figure}

In Fig.~\ref{sfig:torus}, we show the energy spectra of $N_e=8$ and $10$ as a function of the tunneling strength.
For $N_e=8$, with increasing the tunneling strength,
one state in the momentum sector $K=1$ comes down and eventually forms a gapless branch in the low-energy spectrum.
A similar situation occurs also for $N_e=10$. This is the main reason that previous studies ruled out the
possibility of the MR Pfaffian state on the torus\cite{Papic2010}.
Moreover, in Fig.~\ref{sfig:torus}, the energy spectra at small tunneling even fail to develop stable four-fold ground-state degeneracy of the $(331)$ Halperin state.
Therefore, we believe that ED calculations suffer from finite-size effect too strongly to demonstrate the MR Pfaffian physics.

\subsection{Effective Single-component Systems in the Strong-tunneling Limit}\label{app:ED2}
We can reformulate the bilayer FQH system in a new single-particle basis defined by $c_{m,{\rm s}}^\dagger=\frac{1}{\sqrt{2}}(c_{m\uparrow}^\dagger+c_{m\downarrow}^\dagger)$ and $c_{m,{\rm as}}^\dagger=\frac{1}{\sqrt{2}}(c_{m\uparrow}^\dagger-c_{m\downarrow}^\dagger)$, where $c_{m,{\rm s}}^\dagger$ ($c_{m,{\rm as}}^\dagger$) creates an electron in the symmetric (antisymmetric) orbital $m$ between two layers. In this picture, the tunneling term becomes diagonal as $-t_{\perp}\sum_{m=0}^{N_s-1}(c_{m,{\rm s}}^\dagger c_{m,{\rm s}}-c_{m,{\rm as}}^\dagger c_{m,{\rm as}})$. Therefore, in the strong-tunneling limit $t_{\perp}\rightarrow\infty$, the degrees of freedom in the antisymmetric basis are frozen, thus we can view the bilayer system
as an ``effective'' single-component system in the symmetric basis with an average interaction $\frac{1}{2}(H_{{\rm intralayer}}+H_{{\rm interlayer}})$.
This greatly simplifies the problem and makes $N_e=12$ and $16$ accessible in ED.

\begin{figure}[t]
 \begin{minipage}{0.99\linewidth}
 \centering
 \includegraphics[width=0.65\textwidth]{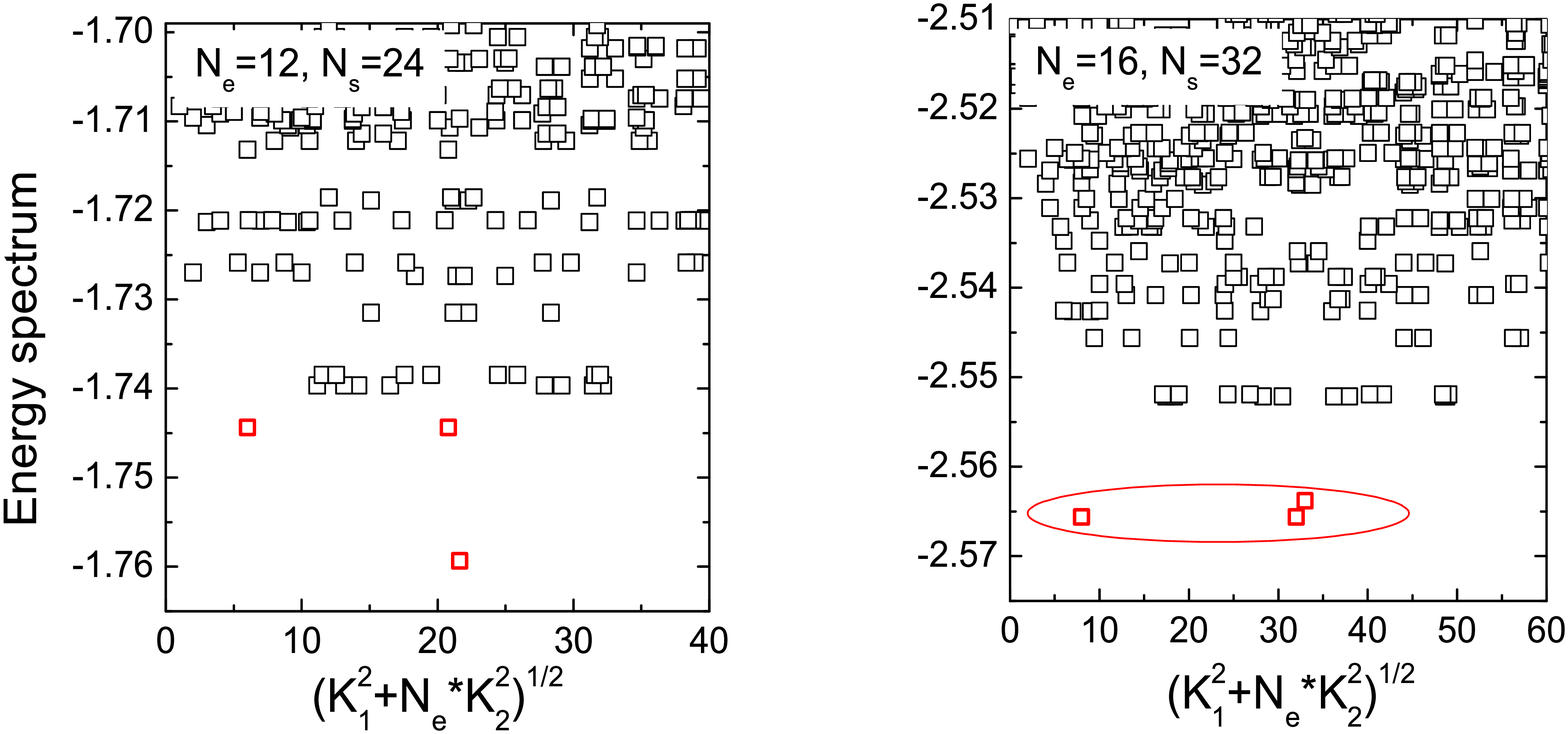}
 \end{minipage}
 \caption{Energy spectra in the strong-tunneling limit for (left) $N_e=12$ and (right) $N_e=16$ on the square torus obtained by ED. Energy eigenstates are labeled by a
 $\sqrt{K_1^2+N_eK_2^2}$ (in unit of $2\pi/N_s$), where $(K_1,K_2)$ is two-dimensional momentum (see Appendix Sec.~\ref{app:method}).
 The three-fold ground-state degeneracy in $(0,\pi),(\pi,0),(\pi,\pi)$ are labeled by red squares.
 All calculations are performed at layer distance $d=3.0$ and layer width $w=1.5$.
 } \label{sfig:torusss}
\end{figure}

In Figs.~\ref{sfig:torusss}, we show the energy spectra of $N_e=12$ and $16$ in this strong-tunneling limit on the torus obtained by ED.
There are three ground states in $(K_1,K_2)=(0,\pi),(\pi,0),(\pi,\pi)$.
We have one remark on the ED calculation. In the strong-tunneling limit, since the system is effectively single-component,
there is an exact particle-hole symmetry in the half-filled lowest Landau level
(while this symmetry is explicitly broken in weak or intermediate tunneling regime of bilayer FQH systems).
An conventional view is that, there should be another copy of the MR Pfaffian degeneracy in the energy spectrum
contributed by the particle-hole conjugate of the MR Pfaffian state, i.e., the anti-Pfaffian state\cite{Peterson2008,HaoWang2009}.
For finite-size systems, two copies (and states in each copy) are split and their eigenstates are symmetric and antisymmetric linear combinations of the MR Pfaffian and anti-Pfaffian states. Apparently, Fig.~\ref{sfig:torusss} shows that
splitting between the symmetric and antisymmetric combinations is not negligible,
indicating that the ED calculation still suffers from strong finite-size effect in the strong-tunneling regime.
However, we should emphasize that our main conclusion focuses on the intermediate-tunneling regime based on DMRG.

\section{Experimental Setup}\label{app:exp}
\subsection{Related Experimental Parameters}
In this section, we briefly review some details of quantum Hall experiments in double quantum well and single wide quantum well systems. In these systems, several physical quantities are tunable in experiments, including the interlayer separation $d$ (in unit of $\ell$), interlayer tunneling strength $t_{\perp}$ (in unit of $e^2/\ell$), and the layer width of a single quantum well $w$ (in unit of $\ell$). Different samples can be constructed with different values of $d$ and $w$. Tunneling strength is determined by the height of the potential barrier between two layers in double quantum well systems or the single-particle wavefunction overlap in single wide quantum well systems. Tuning the parameters can be achieved by varying the electron density $\rho$, which leads to the change of the effective $\ell$ at a fixed filling $\nu$ via the relation $\rho=\nu/2\pi \ell^2$. This allows $d/\ell$, $w/\ell$ and $t_{\perp}/ (e^2/\ell)$ to be tuned continuously in a single sample.

To illustrate the typical parameter range that can be accessed, we show the parameters in several experiments at $\nu_T=1/2$, as shown in Table~\ref{Exp:para}. In double quantum well systems \cite{Eisenstein1992}, it is possible to vary $d$ in the range $2 \sim 4$, while the interlayer tunneling $t_{\perp}$ can be suppressed to $\sim0.01$. The width of individual layers in this case is less than $d$. On the other hand, in wide quantum wells, the effective layer distance can be varied from $2.0$ to $8.0$ \cite{Shabani2013}, and the tunneling strength $t_{\perp}$ typically varies between $0.0$ and $0.2$. For systems where FQH can be observed, the estimated layer width $w$ is typically much smaller than $d$.

\begin{table} \label{Exp:para}
\begin{tabular}{c | c c c  }
\hline\hline
 Experiment \kern10pt &  $d/\ell$ \kern10pt & $t_{\perp}/(e^2/\ell)$ \kern10pt & $w/\ell$ \kern10pt   \\ \hline
Suen \textit{et al.},  Ref.~\onlinecite{Suen1992,Suen1994} & 4.5 - 7.0 & 0.04-0.10 & 2.4-2.5  \\
\hline
Eisenstein \textit{et al.}, Ref.~\onlinecite{Eisenstein1992} & 2.4 -3.6 & 0.01 & 1.8 \\
\hline
Shabani \textit{et al.},  Ref.~\onlinecite{Shabani2013} & 5.0-8.0 & 0.05-0.16 & 2.8-3.2 \\
\hline\hline
\end{tabular}
\caption{Typical parameter values for several existing quantum Hall experiments at $\nu_T=1/2$.
The estimation of layer separation $d$ (in unit of magnetic length $\ell$), quantum well layer width $w$ (in unit of $\ell$), and
tunneling strength $t_{\perp}$ [in unit of Coulomb energy $e^2/\ell$] are taken from
self-consistent calculations \cite{Shabani2013}.
}
\end{table}

\subsection{Related Experimental Measurements}
\begin{figure}[!htb]
 \begin{minipage}{0.98\linewidth}
 \centering
 \includegraphics[width=0.32\textwidth]{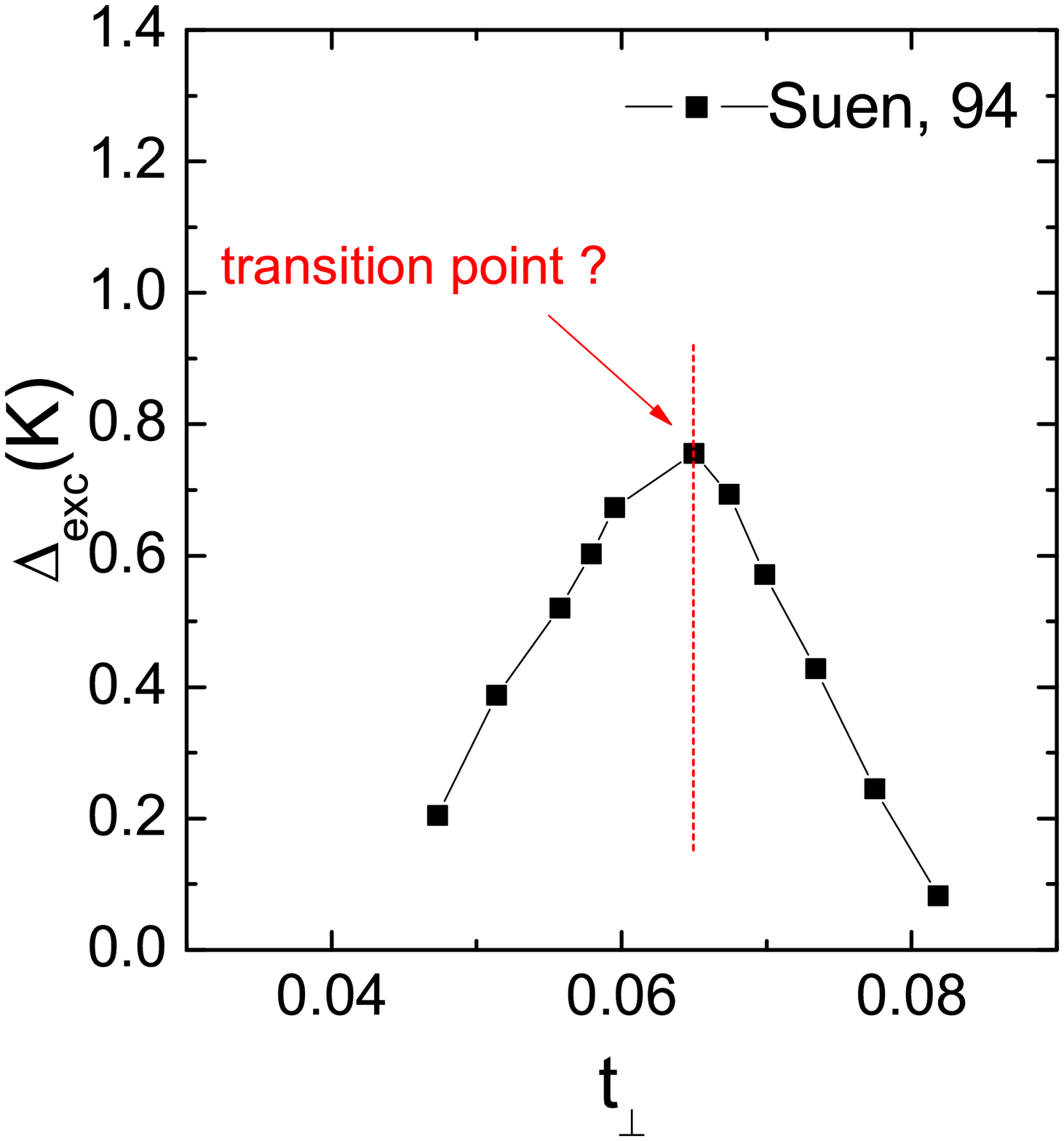}
 \includegraphics[width=0.4\textwidth]{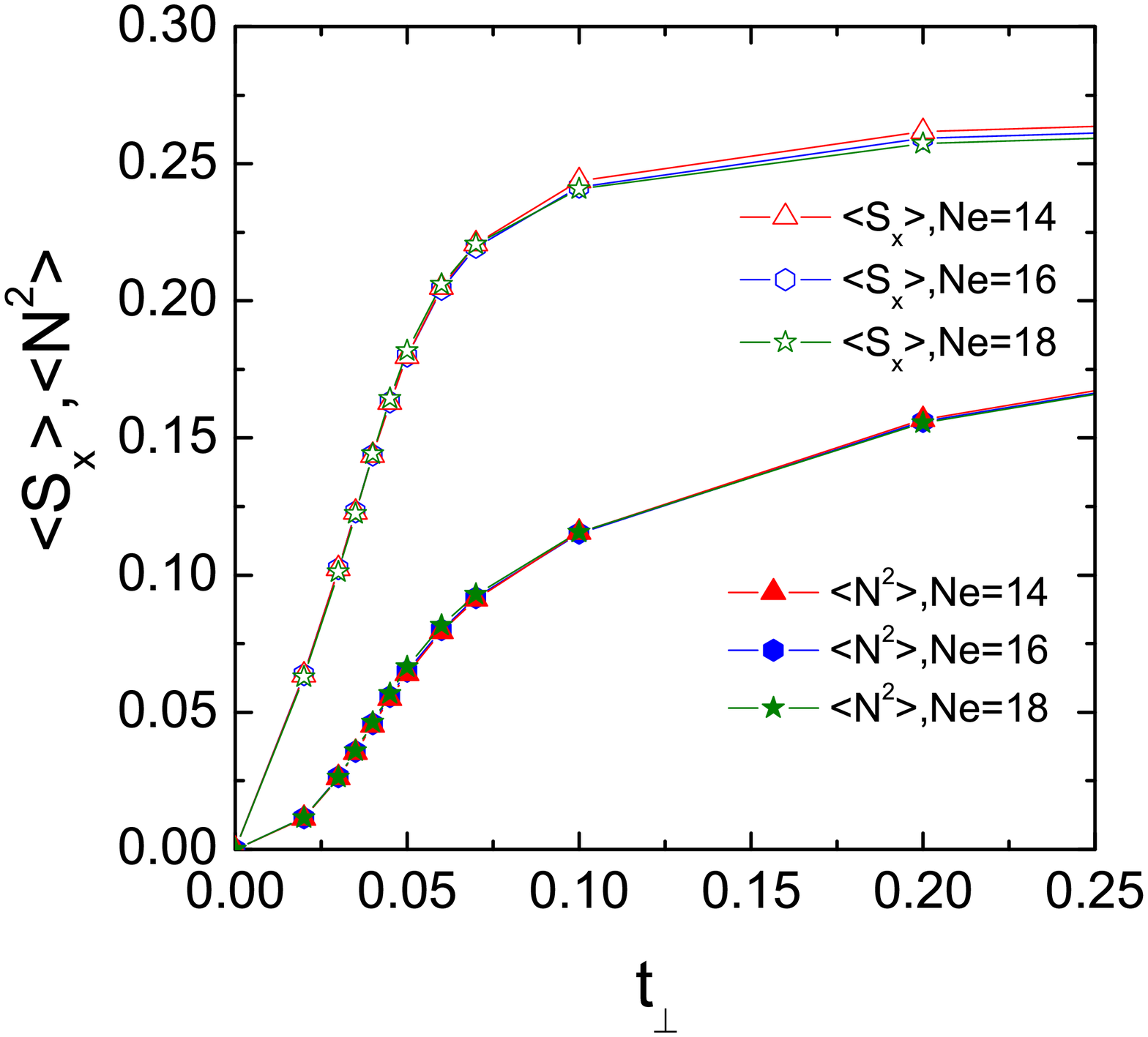}
 \end{minipage}
 \caption{(left) The experimental observed quasiparticle excitation gaps $\Delta_{\rm exc}$ (in unit of Kelvin) versus the tunneling strength $t_{\perp}$.
 The data is obtained from Ref.~\onlinecite{Suen1994}.
 (right) The tunneling current and particle number fluctuation 
of each layer as a function of $t_{\perp}$ with layer width $w=1.5$ and layer separation $d=3.0$, which can be used to distinguish a two-component state from a single-component state.
 } \label{measure}
\end{figure}

We briefly discuss the experimentally related observations.
First of all, we emphasize that the evolution of excitation gap $\Delta_{\rm exc}$ with tunneling strength $t_{\perp}$
is qualitatively consistent with the experimental observations.
As shown in Fig. \ref{measure}(left), by tuning the effective tunneling strength $t_\perp$,
the measured quasiparticle excitation gap $\Delta_{\rm exc}$ develops an upward cusp behavior.
Compared with our results in Fig. \ref{transition}(c) in the main text,
we conclude that the presence of a maximum in excitation gap in the existing experiment \cite{Suen1994}
is a direct signal of the phase transition from the $(331)$ Halperin state to the MR Pfaffian state.

Next, we propose several methods to distinguish the $(331)$ Halperin state and the MR Pfaffian state in experiments.
First, the $(331)$ Halperin state has a quantized drag Hall conductance $\sigma^{xy}_{\rm drag}= -\frac{e^2}{8h}$, but the drag Hall conductance of the MR Pfaffian state is not quantized. Therefore, we can measure $\sigma^{xy}_{\rm drag}$ to distinguish them. If there are
separate electric contacts in two different layers, and the
electric current $\delta I_{\uparrow}$ is forced to flow in the top layer (also called the
driving layer), a measurable voltage
drop $\delta V_{\downarrow}$ will be induced in the bottom layer (called the drag layer).
On the other hand, one can also measure the tunneling current $j_t \propto \langle S_x\rangle =\frac{1}{N_{\phi}}\sum_i \langle c^\dagger_{i,\uparrow} c_{i,\downarrow}+{\rm h.c.}\rangle$ to distinguish these two states.
For the $(331)$ Halperin state, the tunneling current between two layers should be small and
sensitive to the change of the tunneling strength, as shown in Fig.~\ref{measure}(right).
On the contrary, the tunneling current keeps finite and almost does not change
for the MR Pfaffian state.
Moreover, the $(331)$ Halperin state and the MR Pfaffian state can also be distinguished by measuring the particle number fluctuation
$\langle N^2\rangle =\langle \delta N^2\rangle/N_e =\langle (N_{\uparrow}-N_e/2)^2\rangle/N_e=\langle (N_{\downarrow}-N_e/2)^2\rangle/N_e$ in each layer.
Since electrons are confined in two layers in the $(331)$ Halperin state, the particle number fluctuation should be strongly suppressed.
But a large particle number fluctuation is expected for the MR Pfaffian state due to the tunneling effect, as shown in Fig.~\ref{measure}(right).

\end{appendices}

\end{document}